\documentclass[prd,eqsecnum,floatfix]{revtex4}%
\usepackage{amsmath}
\usepackage{graphicx}
\usepackage{amsfonts}
\usepackage{amssymb}%
\providecommand{\U}[1]{\protect\rule{.1in}{.1in}}
\newcommand{\BOX}{\hbox {$\sqcap$ \kern -1em $\sqcup$}}

\newcommand{\be}{\begin{equation}}
\newcommand{\ee}{\end{equation}}
\newcommand{\ba}{\begin{eqnarray}}
\newcommand{\ea}{\end{eqnarray}}
\newcommand{\ban}{\begin{eqnarray*}}
\newcommand{\bea}{\begin{eqnarray}}
\newcommand{\eea}{\end{eqnarray}}
\newcommand{\ean}{\end{eqnarray*}}
\newcommand{\barr}{\begin{array}}
\newcommand{\earr}{\end{array}}

\begin{document}
\title{Towards a microscopic construction of flavour vacua from a space-time foam model}
\author{Nick E. Mavromatos and Sarben Sarkar}
\affiliation{King's College London, University of London, Department of Physics, Strand,
London WC2R 2LS, U.K.}

\begin{abstract}
{\small The effect on flavour oscillations of simple expanding background
space-times, motivated by some D-particle foam models, is calculated for a
toy-model of bosons with flavour degrees of freedom. The presence of
D-particle defects in the space-time, which can interact non trivially (via
\textquotedblleft particle capture\textquotedblright) with flavoured particles
in a flavour non-preserving way, generates mixing in the effective field
theory of low-energy string excitations. Moreover, the recoil of the
D-particle defect during the capture/scattering process implies Lorentz
violation, which however may be averaged to zero in isotropic D-particle
populations, but implies non-trivial effects in correlators. Both features
imply that the flavoured mixed state sees a non-trivial flavour (Fock-space)
vacuum of a type introduced earlier by Blasone and Vitiello in a generic
context of theories with mixing. We discuss the orthogonality of the flavour
vacua to the usual Fock vacua and the effect on flavour oscillations in these
backgrounds. Furthermore we analyse the equation of state of the Flavour
vacuum, and find that, for slow expansion rates induced by D particle recoil,
it is equivalent to that of a cosmological constant. Some estimates of these
novel non-perturbative contributions to the vacuum energy are made. The
contribution vanishes if the mass difference and the mixing angle of the
flavoured states vanish.}

\end{abstract}
\maketitle

%\date{\today}

\section{Introduction and Motivation}

It has been suggested by Blasone and Vitiello and
collaborators~\cite{vitiello} that in quantum field theoretic systems with
\emph{mixing}, such as neutrinos with flavour mixing, for instance, the
\textquotedblleft flavour\textquotedblright\ eigenstates can be expressed as a
condensate in the Fock space of the mass quanta. Furthermore, it has been
shown recently that there is a "vacuum energy" associated with this condensate
which would be a contribution to the dark energy. In contrast to the standard
mass-eigenstate formalism, where the vacuum expectation value (vev) of the
stress tensor vanishes, when one calculates the vev of the stress tensor with
respect the flavour vacuum, the result is non-trivial, indicating spontaneous
breaking of Lorentz-symmetry. In fact, computing such vevs for the $T_{00}$
and $T_{ii}$ components of the stress tensor in Robertson-Walker space times,
the authors of \cite{capolupo} have conjectured that there is a non-trivial,
and non-perturbative, dark energy contribution to the vacuum, and an equation
of state $p=w\rho$ with $w<0$ close to $-1$. These are features that are
compatible with recent observations.

In view of the breaking of Lorentz-symmetry by the flavour vacuum,
characterized by a non-zero vev of the stress tensor, we will examine in this
context a microscopic model discussed by Mavromatos and Sarkar~\cite{mavsark}
where such a feature is accommodated within a specific space-time foam model
in the framework of modern brane/string theory. The foam model involves brane
worlds which are embedded in higher-dimensional bulk space-times, and are
punctured by D0-brane (point-like defects)~\cite{emw}. The interaction of
generic stringy matter with such defects involves the discontinuous process of
splitting of the matter string by the defect, and subsequent joining, ignoring
for the moment constraints from conservation of specific quantum numbers .
This yields Lorentz-violating metric distortions~\cite{recoil,recoil2},
$g_{\mu\nu}\propto u_{\mu}u_{\nu}$, with $u_{\mu}$ the recoil velocity of the
localised D-particle (D0) defect, due to the \textquotedblleft
topologically-non-trivial\textquotedblright\ interaction with the stringy
matter as well as an induced local expansion. In a situation where one
encounters a statistical population of D0-defects, one may have an average
$\langle\langle u_{i}\rangle\rangle=0$ (i.e. isotropic models for D-particle
foam). Hence, since $u_{\mu}u^{\mu}=1$ for a velocity four-vector, one has a
preferred time-like direction in space time in this case, leading to induced
Lorentz-symmetry breaking from non-trivial higher order correlators of
$u_{\mu}$. Such a mechanism for Lorentz-symmetry violation is testable since
it leads to a quite distinct signature for CPT violation~\cite{greenberg} in
meson factories~\cite{mavsark2}, associated with a modification of the
Einstein-Podolsky-Rosen(EPR)-type particle correlations. Electric charge
conservation implies that charged states cannot undergo such interactions.
Moreover "flavour" need not be conserved in the string splitting process. If
D-particle foam contributes to the mixing of flavours then it only operates on
the electrically-neutral neutrinos (or mesons). The above-mentioned
Lorentz-violating \textquotedblleft flavour-vacua\textquotedblright\ appear to
be the physically correct ones to characterize the interaction of electrically
neutral flavour matter (which for brevity we will refer to as neutrinos) with
D0 defects. \ \ \ 

It should be pointed out that in the work of \cite{vitiello,capolupo} the
flavour vacuum formalism applies in general to \emph{all} theories with
mixing. In the above specific framework of stringy models of D-particle foam,
only the \emph{neutral }flavour particles are subject to mixing and contribute
to the dark energy. Dark energy, when interpreted as a cosmological constant,
gives a de Sitter space-time with an associated horizon. Hence, it is
interesting for self-consistency to re-examine and reinterpret the vacuum
oscillation phenomena in simple space-times with horizons. In our case the
latter are interpreted as the metric distortions due to D-particle \ recoil
and will be shown to be most important for low energy neutrinos. This is to be
contrasted with the approach of \cite{capolupo} where an \emph{a priori} flat
space-time formalism was used to discuss the derivation of the dark energy
contribution due to mixing. For simplicity we consider a two dimensional
space-time metric as a physically-relevant toy model to illustrate the basic
features of the approach. The restriction in space-time dimensionality is
plausible if the interaction and recoil of neutrino matter with the D0 brane
defects (D-particles), is mainly along the (spatial) direction of the incident
neutrino beam.

The paper is organised as follows. We consider the recoil of D particles after
the capture and subsequent emission of stringy matter in terms of suitable
vertex operators~\cite{recoil,recoil2} within a logarithmic field theory
\cite{gurarie}. By considering an ensemble of defects we model the metric
induced by the recoil as a simple two-dimensional space-time with a
horizon~\cite{birrell}. Finite systems on a lattice with periodic boundary
conditions will be considered and the thermodynamic limit will be taken
subsequently. This allows a rigorous treatment of the orthogonality ( and also
of unitary inequivalence~\cite{vitiello,ji}) of the flavour and mass vacua.
The effect of the space-time expansion on neutrino oscillations is evaluated.
In order to check if a dark energy interpretation of the equation of state is
allowed, the equation of state of the flavour vacuum (after expansion) is
calculated. Given the brane/string motivation for the model, a novel form of
normal-ordering is introduced which demonstrates a contribution to dark
energy. It is important to notice that the type of equation of state obtained
by the above normal ordering procedure turns out to be formally independent of
the initially assumed form of the scale factor of the Robertson-Walker
background space time, thereby allowing for a self-consistent treatment of the
back reaction \emph{non-perturbative} effects of the flavour vacuum onto the
space time. Some technical details of our approach are outlined in an Appendix.

\section{String inspired toy model}

\bigskip The discovery of new solitonic structures in superstring theory has
dramatically changed the understanding of target space structure. These new
non-perturbative objects are known as D-branes and their inclusion leads to a
scattering picture of space-time fluctuations. The study of D-brane dynamics
has been made possible by Polchinski's realisation that such solitonic string
backgrounds can be described in a conformally invariant way in terms of world
sheets with boundaries \cite{polchinski}. On these boundaries Dirichlet
boundary conditions for the collective target-space coordinates of the soliton
are imposed. The properties of D-branes in curved backgrounds is not fully
developed and so it is necessary to be guided by approximate calculations. In
the interaction of low energy matter (e.g. a closed string) with a heavy D
particle (i.e. embedded in a $\left(  d+1\right)  $-dimensional space-time the
D-particle recoils. Such recoil fluctuations of D0 branes can be described by
an effective stochastic space-time. From our point of view there are two parts
to this interaction process: precursor and aftermath of the capture of a
string by a D-particle. Firstly the stringy matter has to approach the
D-particle. In a superstring approach by considering the annulus amplitude for
an open string fluctuation between a D$p$ and D$p^{\prime}$ brane which have a
non-zero relative velocity it can be shown that there is an attractive
potential between them \cite{emw}. By specialising to $p=1$ and $p^{\prime}=0$
we can give arguments as to why the low momenta strings are more likely to be
captured by the D-particle. Secondly we have the capture and the re-emission
of a string by the D-particle. This involves a change in the background of the
string and as such breaks conformal invariance. It is not known how to
describe the process of capture for short times. It is the capture and
re-emission process which can \ be responsible \ for a change in the flavour
of the matter string. For long times after the capture and emission process a
semi-classical approach is expected to be valid \cite{gravanis}. In the wake
of the capture process the vertex operator of the D-particle is described by
an impulse approximation. The breaking of conformal invariance for large times
can still be dealt with in terms of conformal data but with a set of recoil
vertex operators which satisfy a logarithmic conformal algebra. Such an
impulse/recoil event results in general in a time-dependent induced metric
distortion and vacuum energy.\ It is such a metric distortion that is the
motivation for considering the simple model that we will consider.

The time scale of the D-particle matter interaction can be estimated from
perturbative string theory. In the adiabatic approximation for the relative
speed $u$ between a p-brane and a D-particle separated by a distance $r$, it
is possible to estimate the potential energy $V\left(  r,u\right)  $ using
perturbative string theory \cite{emw}. Typically \ the short range attractive
part of the potential $\mathcal{V}$ is given by
\begin{equation}
\mathcal{V}\left(  r,u\right)  \sim-\frac{2\pi\ell_{s}^{2}u^{2}}{r^{3}},
\label{pot}%
\end{equation}
with $\ell_{s}$ being the string scale and is the only relevant piece in a
superstring theory. For the case of flavour oscillations with sharp momentum
$k$ (whose associated mass eigenstates have masses $m^{\left(  1\right)  }$
and $m^{\left(  2\right)  }$ with $m^{\left(  1\right)  }>m^{\left(  2\right)
}$ in our convention$\,$) the time scale $\tau$ for the flavour changing
capture and subsequent release of a matter particle by a heavy D-particle
(when $\hbar=c=1$) can consequently be estimated by%
\[
\tau\sim\frac{1}{\Delta\mathcal{V}}%
\]
where $\Delta\mathcal{V}=\frac{2\pi}{\ell_{s}}\left\vert \frac{k^{2}%
}{m^{\left(  2\right)  2}}-\frac{k^{2}}{m^{\left(  1\right)  2}}\right\vert
=\frac{2\pi k^{2}\Delta m^{2}}{\ell_{s}m^{\left(  1\right)  2}m^{\left(
2\right)  2}}$and $\Delta m^{2}\equiv m^{\left(  1\right)  2}-m^{\left(
2\right)  2}$. Hence we deduce that qualitatively
\begin{equation}
\tau\sim\frac{\ell_{s}}{2\pi}\frac{m^{\left(  1\right)  2}m^{\left(  2\right)
2}}{\Delta m^{2}}\frac{1}{k^{2}}. \label{interactiontime}%
\end{equation}
We shall then assume that the capture process is enhanced for low $k$. We can
give an estimate for this time scale on considering a typical situation
$k=1$GeV, $m^{\left(  1\right)  }\sim m^{\left(  2\right)  }\sim10^{-1}$eV and
$\Delta m^{2}\sim10^{-5}$eV$^{2}$; we find $\tau\sim10^{-60}\,$sec which is
effectively instantaneous.

\begin{figure}[th]
\centering
\includegraphics[width=7.5cm]{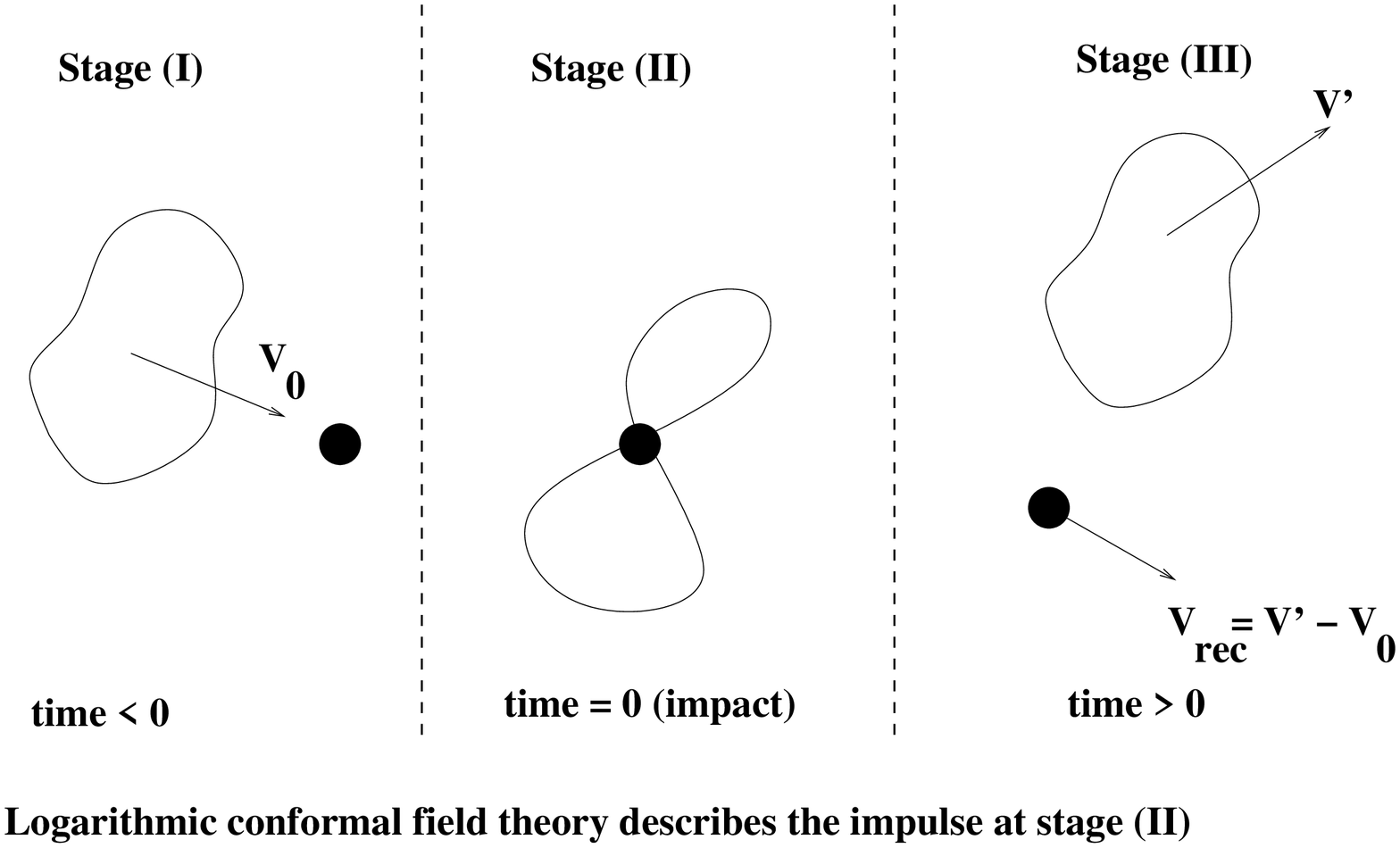} \hfill
\includegraphics[width=7.5cm]{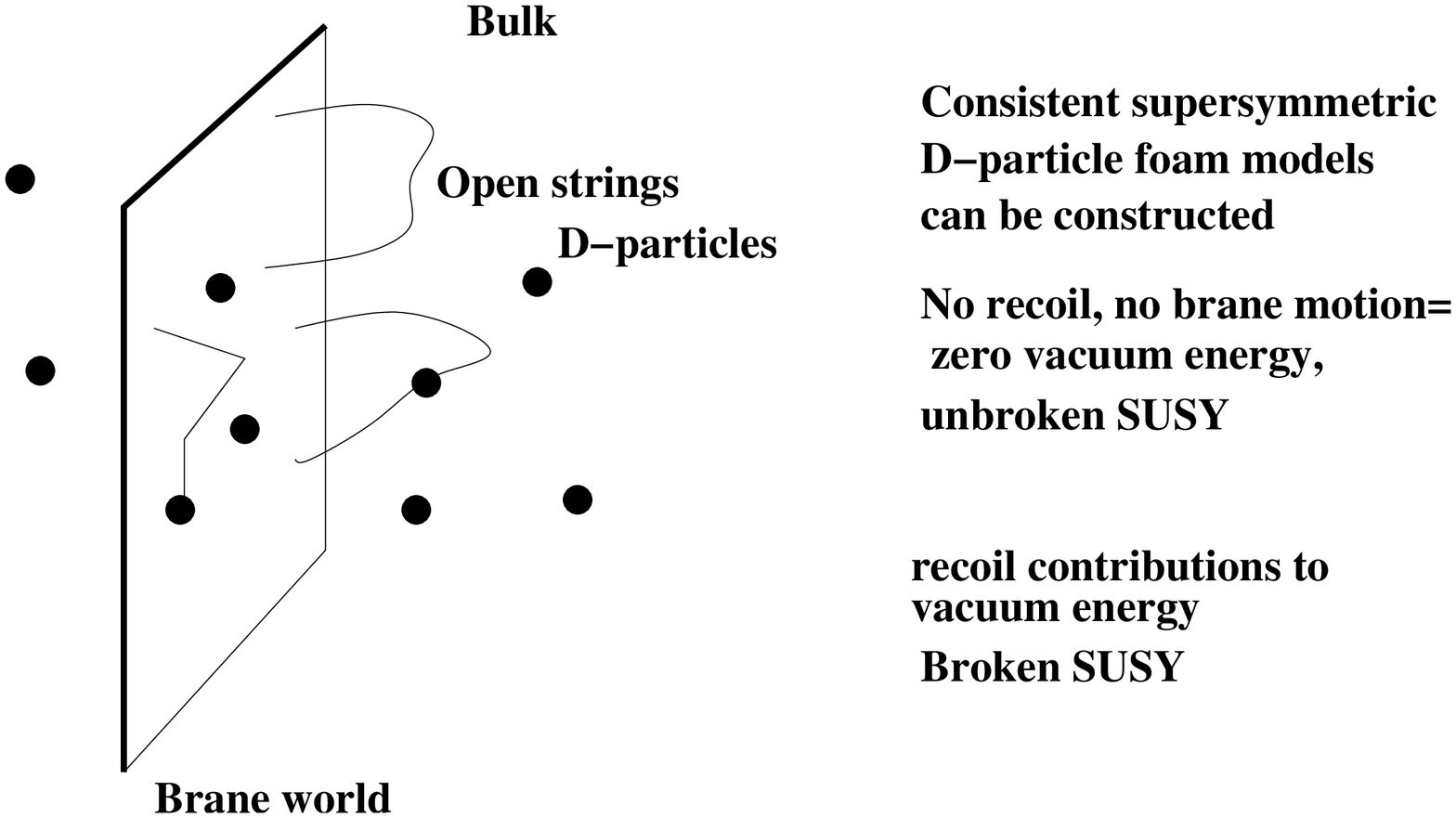} \caption{Schematic
representation of the capture/recoil process of a string state by a D-particle
defect for closed (left) and open (right) string states, in the presence of
D-brane world. The presence of a D-brane is essential due to gauge flux
conservation, since an isolated D-particle cannot exist. The intermediate
composite state at $t=0$, which has a life time within the stringy uncertainty
time interval $\delta t $, of the order of the string length, and is described
by world-sheet logarithmic conformal field theory, is responsible for the
distortion of the surrounding space time during the scattering, and
subsequently leads to Finsler-type~\cite{finsler} induced metrics (depending
on both coordinates and momenta of the string state) and modified dispersion
relations for the string propagation.}%
\label{fig:recoil}%
\end{figure}

We, for definiteness, consider a D3 brane world with D-particles in the bulk
scattering off the brane. The recoil of a D-particle on scattering with a
closed (or open) string (see fig.~\ref{fig:recoil}) leads to a world sheet
deformation given by the semi-classical vertex operator $V$
\begin{equation}
V=\int_{\partial\Sigma}G_{ij}y^{i}\left(  X^{0}\right)  \partial_{n}X^{j}
\label{vertex}%
\end{equation}
where $G_{ij}$ is given by the spatial part of the space-time metric,
$\partial\Sigma$ the world sheet boundary, $\partial_{n}$ is a derivative on
the worldsheet normal to the boundary, $X^{j}$ is a spatial target space field
satisfying Dirichlet boundary conditions and $X^{0}$ is the timelike target
space field satisfying Neumann conditions (in the standard string $\sigma$
model) and $y^{i}(X^{0})$ is the classical D-particle trajectory which, with a
suitable choice of co-ordinates, vanishes before the time $t_{0}$ of the
impulse. Let \ $k_{1}\left(  k_{2}\right)  $ be the momentum of the
propagating closed-string state before (after) the recoil, $y_{i}$ be the
spatial collective coordinates of the D particle, $g_{s}\left(  <1\right)  $
is the string coupling and $M_{s}$ is the string scale and $\sqrt
{\alpha^{\prime}}\varepsilon^{-1}$($\alpha^{\prime}$ being the Regge slope) be
identified with the target Minkowski time $X^{0}$ for $X^{0}\gg0$ after the
collision For an initially Minkowski space-time background, we
have~\cite{recoil,recoil2}:
\begin{equation}
y_{i}(X^{0}(z))=\delta_{ij}y^{j}(X^{0}(z))=\mathcal{C}_{i}\left(  z\right)
+\mathcal{D}_{i}\left(  z\right)
\end{equation}
where $z=e^{-i\left(  \sigma+i\tau\right)  }$, $\left(  \sigma,\tau\right)  $
being the standard world sheet co-ordinates, and the operators $\mathcal{C}%
_{i}\left(  z\right)  $ and $\emph{D}_{i}\left(  z\right)  $ are given (in
terms of the regularised Heavyside function $\Theta_{\varepsilon}$ and
$u_{i}\equiv\frac{g_{s}}{M_{s}}\left(  k_{1}-k_{2}\right)  _{i}\;$) by
\begin{equation}
\mathcal{C}_{i}\left(  z\right)  =\varepsilon y_{i}\Theta_{\varepsilon}\left(
X^{0}\left(  z\right)  /\sqrt{\alpha^{\prime}}\right)  \label{logconf1}%
\end{equation}
and
\begin{equation}
\mathcal{D}_{i}\left(  z\right)  =u_{i}X^{0}\left(  z\right)  \Theta
_{\varepsilon}\left(  X^{0}\left(  z\right)  /\sqrt{\alpha^{\prime}}\right)  .
\label{logconf2}%
\end{equation}
We define $\Theta_{\varepsilon}$ through
\begin{equation}
\Theta_{\varepsilon}\left(  X^{0}\left(  z\right)  /\sqrt{\alpha^{\prime}%
}\right)  \equiv\frac{1}{2\pi i}\int_{-\infty}^{\infty}\frac{dq}%
{q-i\varepsilon}e^{iqX^{0}\left(  z\right)  /\sqrt{\alpha^{\prime}}%
},\,\varepsilon\rightarrow0+. \label{theta}%
\end{equation}
\bigskip We shall from now on choose units with $\alpha^{\prime}=1$. These
vertex operators have been calculated for non-relativistic branes where
$u_{i}$ is small. Consider now the operators $\mathcal{C}\equiv\mathcal{C}%
_{i}\partial_{n}X^{i}$, and $\mathcal{D}\equiv\mathcal{D}_{i}\partial_{n}%
X^{i}$. By calculating the correlators
\begin{align}
\left\langle \mathcal{C}\left(  z\right)  \mathcal{C}\left(  w\right)
\right\rangle  &  \sim O\left(  \varepsilon^{2}\right)  ,\label{logconf3}\\
\left\langle \mathcal{C}\left(  z\right)  \mathcal{D}\left(  w\right)
\right\rangle  &  \sim\frac{1}{\left\vert z-w\right\vert ^{2}},
\label{logconf4}%
\end{align}
and
\begin{equation}
\left\langle \mathcal{D}\left(  z\right)  \mathcal{D}\left(  w\right)
\right\rangle \sim\frac{\log\frac{\left\vert z-w\right\vert }{L}}{^{\left\vert
z-w\right\vert ^{2}}} \label{logconf5}%
\end{equation}
with respect to the measure of the free bosonic string action we see that we
have a logarithmic conformal field theory \cite{gurarie}, \emph{ provided}
that~\cite{recoil} $\varepsilon^{2}\sim\left(  \log\frac{L}{a}\right)  ^{-2}$,
where $L$ is the diameter of the world sheet disc and $a$ is an ultraviolet
world-sheet cut-off. Hence $\varepsilon$ is non-zero for finite $L$. \ 

By considering the operator product of these operators with the stress energy
tensor we find an anomalous dimension $\Delta=\frac{\varepsilon^{2}}{2}$. In
order to restore conformal invariance we need Liouville dressing with a
dilaton field. For simplicity if we restrict the slow recoil to the $x$
direction (on the brane) we obtain the vertex operator $V_{L}^{\left(
x\right)  }$
\begin{equation}
V_{L}^{\left(  x\right)  }=\int_{\Sigma}d^{2}z\,e^{\alpha_{0x}\phi}%
\partial_{\alpha}\left(  u_{x}X^{0}\left(  z\right)  \Theta_{\varepsilon
}\left(  X^{0}\left(  z\right)  \right)  \partial^{_{\alpha}}X^{x}\left(
z\right)  \right)  \label{dressed vertex}%
\end{equation}
where $\phi$ is the dilaton field. From considerations associated with
conformal invariance of the bulk world sheet of the dressed theory
\cite{distler}
\begin{equation}
\alpha_{0x}=-\frac{Q}{2}+\sqrt{\frac{Q^{2}}{4}+\frac{\varepsilon^{2}}{2}}
\label{conformaldim1}%
\end{equation}
where $Q$ is the central charge deficit of the $\sigma$-model describing the
stringy excitations of the recoiling D-particle. We remind the reader that the
restoration of the conformal invariance by the Liouville $\phi$ system
implies
\begin{equation}
c_{\phi}=1+3Q^{2}. \label{conformaldim2}%
\end{equation}
where $c_{\phi}$ is the central charge of the Liouville part of the $\sigma
$-model. As discussed in detail in \cite{recoil,recoil2}, one may argue that,
in our case, $Q^{2}$ is $\mathcal{O}\left(  \varepsilon^{4}\right)  $,
assuming a critical bulk space-time dimension (c.f. fig.~\ref{fig:recoil});
hence, to leading order in $\varepsilon\rightarrow0^{+}$, its contributions
can be ignored in comparison to the anomalous dimension term $\varepsilon
^{2}/2$ in (\ref{conformaldim1}). Thus, we have $\alpha_{0x}\sim
\frac{\varepsilon}{\sqrt{2}}$ (for $u\ll1$) and from (\ref{dressed vertex}) we
obtain:
\begin{equation}
V_{L}^{\left(  x\right)  }\approx\int_{\Sigma}e^{\frac{\varepsilon}{\sqrt{2}%
}\phi}\partial_{\alpha}X^{0}\left(  z\right)  \partial^{\alpha}X^{x}\left(
z\right)  u_{x}\Theta_{\varepsilon}\left(  X^{0}\left(  z\right)  \right)
\,+\dots\label{dressed vertex2}%
\end{equation}
where the $\dots$ denote terms that do not contribute to physical correlators.
Such terms are proportional to either $X^{0}\delta(X^{0})$ (obtained by the
action of $\partial_{\alpha}$ on the Heavyside operator $\Theta_{\varepsilon
}(X^{0})$), or $\partial_{\alpha}\partial^{\alpha}X^{i}$ (which vanish in the
conformal gauge on the world sheet).

In the recoil case the string is slightly \emph{supercritical}, i.e. $Q^{2}%
>0$. In such a case, the zero mode $\phi_{\left(  0\right)  }$ of $\phi$ can
be interpreted as a second time in addition to $X^{0}$ \cite{antoniadis}. From
(\ref{dressed vertex2}) we then observe that the effect of $V_{L}^{\left(
x\right)  }$ in the action can be reinterpreted as an induced target space
metric in the extended $D+1$-dimensional space-time $(\phi_{\left(  0\right)
},X^{0},X^{i})$
\begin{equation}
G_{\phi\phi}=1,~G_{xx}=-1,~G_{00}=-1,~G_{0x}=e^{\frac{\varepsilon}{\sqrt{2}%
}\phi}u_{x}\Theta_{\varepsilon}\left(  X^{0}\right)  \label{metricrecoil}%
\end{equation}
where, following \cite{gravanis}, we have assumed euclideanised $X^{0}$, as
required for convergence of the world-sheet path integral.

At first sight, it appears that this metric is characterised by two times,
$\phi_{\left(  0\right)  }$ and $X^{0}$. However, it has been
argued~\cite{gravanis} that the open string excitations on the brane world are
such that they live on a $D$-dimensional hyper-surface of the extended
$D+1$-dimensional two-time space-time; the Liouville mode is proportional to
$\sum_{\mu=0}^{3}u_{\mu}X^{\mu}$,with $u_{\mu}$ the four-velocity on the D3
brane for the recoiling D-particle and $u_{\mu}$ can be written as
$\gamma(1,\vec{u}_{i})$, where $\gamma\left(  =1/\sqrt{1-u_{i}^{2}}\right)  $
is the Lorentz factor .

We shall demonstrate this structure in an explicit example in which the
three-brane world is obtained by compactification on magnetized tori,
characterised by a magnetic field with intensity $H$. This magnetic field is
not a real one, but a \textquotedblleft statistical\textquotedblright\ field
characterising the compactification. An open string excitation coupled to such
a magnetic field, has its ends attached to the compactified three brane world.
An encounter of the string with a D-particle is described by the addition of
the following deformation to a free open-string $\sigma$-model action:
\begin{equation}
\int_{\partial\Sigma}H\Theta_{\varepsilon}(u_{\mu}X^{\mu})X^{4}\partial^{\tau
}X^{5} \label{deformsudden}%
\end{equation}
where $\partial^{\tau}$ denotes tangential derivative on the world-sheet
boundary. This corresponds to the vertex operator for a gauge potential
corresponding to a uniform magnetic field in the compactified toroidal
dimensions $(X^{4},X^{5})$, that appears suddenly at a time $\sum_{\mu=0}%
^{3}u_{\mu}X^{\mu}=0$, in the co-moving frame of the recoiling D-particle,
which has captured the open string excitation. The reader should notice that
the specific calculation takes place in this co-moving frame; the argument of
the Heavyside operator is the rest-frame time of the recoiling massive
D-particle, which in terms of the original coordinates is expressed as
$u_{\mu}X^{\mu}=\gamma(X^{0}+u_{i}X^{i})$.

It can be easily seen~\cite{recoil} that the operator (\ref{deformsudden}) has
anomalous dimension $-\frac{\varepsilon^{2}}{2}u_{\mu}u^{\mu}=-\frac
{\varepsilon^{2}}{2}$, since the four velocity satisfies by definition
$u_{\mu}u^{\mu}=1$, with respect to the background Minkowski metric. This
implies a Liouville dressing similar to (\ref{dressed vertex}), with the
corresponding Liouville anomalous dimension $\alpha\sim\frac{\varepsilon
}{\sqrt{2}}$,
\begin{equation}
\int_{\partial\Sigma}e^{\frac{\varepsilon}{\sqrt{2}}\phi}H\Theta_{\varepsilon
}(u_{\mu}X^{\mu})X^{4}\partial^{\tau}X^{5}. \label{suddenmagn2}%
\end{equation}

As shown in \cite{gravanis} the open superstring excitations that couple to
such a magnetic field will exhibit Zeeman-like supersymmetric mass splittings
between fermions and bosons~ of the form:
\begin{equation}
\delta m^{2}\sim He^{\frac{\varepsilon}{\sqrt{2}}\phi}H\Theta_{\varepsilon
}(u_{\mu}X^{\mu})\Sigma_{45} \label{split}%
\end{equation}
with $\Sigma_{45}$ the spin operator given by $\frac{i}{4}\left[  \gamma
_{4},\gamma_{5}\right]  $ where $\gamma_{\mu}\,\left(  \mu=0,\ldots,5\right)
$ are the Dirac matrices relevant for $6$ dimensional space-time. However
masses need to be time (or equivalently $\varepsilon$) independent for
stability. Representing the Heavyside operator for positive arguments as
$\Theta_{\varepsilon}(u_{\mu}X^{\mu})\sim e^{-\varepsilon u_{\mu}X^{\mu}}$,
$\delta m^{2}$ is independent of the value of $\varepsilon$
if~\cite{gravanis}
\begin{equation}
\frac{\phi_{\left(  0\right)  }}{\sqrt{2}}-\sum_{\mu=0}^{3}u_{\mu}X^{\mu
}=\mathrm{constant}~ \label{liouvtime}%
\end{equation}
where the constant can be taken to zero without loss of generality. Since the
above-mentioned mass splittings can be tuned by adjusting appropriately the
intensity of the \textquotedblleft magnetic\textquotedblright\ field $H$, in a
way independent of the size of the four-dimensional vacuum energy
contribution~\cite{gravanis}, the above example is sufficiently generic and so
of interest for phenomenological applications of superstring theories.

The constraint (\ref{liouvtime}), which is consistent with general coordinate
invariance on the brane world, implies that the dynamics of the open string
excitations can effectively be considered on a $D$-dimensional space-time
(hypersurface), with a \emph{ single} time. Although above we have argued in
favour of (\ref{liouvtime}) in the context of a specific superstring
situation, nevertheless we may assume its validity in more general frameworks
of supercritical strings~\cite{emn}.

From (\ref{liouvtime}) and (\ref{metricrecoil}) we then obtain for the induced
$D$-dimensional target space-time metric (with $X^{0}\equiv t$):
\begin{equation}
ds^{2}=d\phi^{2}-dt^{2}+u_{x}dXdt-(dX)^{2}=2\gamma^{2}(dt+u_{x}dX)^{2}%
-dt^{2}+u_{x}dXdt-(dX)^{2} \label{recmetric}%
\end{equation}
We next consider the case of a statistically significant population of
D-particles on the brane world (c.f. fig.~\ref{fig:recoil}) (\textquotedblleft
D-particle foam\textquotedblright). We may assume a random (Gaussian)
D-particle foam, such that~\cite{mavsark}:
\begin{equation}
\langle\langle u_{x}\rangle\rangle=0~,\qquad\langle\langle u_{x}u_{x}%
\rangle\rangle=\sigma^{2}, \label{gaussian}%
\end{equation}
and we assume slowly moving D-particles, so that cubic or higher powers of the
recoil velocities are ignored. The average of the metric (\ref{recmetric}),
then, yields (upon expanding appropriately the Lorentz factors in powers of
$u_{x}^{2}$):
\begin{equation}
\langle\langle ds^{2}\rangle\rangle=\left(  1+2\sigma^{2}\right)
dt^{2}-(1-2\sigma^{2})dX^{2}+dY^{2}+dZ^{2} \label{metric2b}%
\end{equation}
Notice the existence of \emph{horizons} in this metric, the \textquotedblleft
xx\textquotedblright\ metric component vanishes for $\sigma^{2}=1/2$ (which is
consistent with the fact that the velocities are subluminal $|u_{\mu}|<1$ ).
The metric (\ref{metric2b}) has a \emph{ global} character, as a result of
averaging over populations of D-particles that cross the \emph{ entire }
D-brane world.

We next notice that the quantities $\sigma^{2}$ are in principle time
dependent, denoted by $\sigma^{2}(t)$, since the density of the D-particle
defects may depend on time. This can happen, for instance, if the bulk density
of the D-particles is not uniform. The nature of the time dependence of
$\sigma^{2}(t)$ is therefore dependent on a microscopic model. If the
\textit{density} of the D-particles that cross our brane world (c.f.
fig.~\ref{fig:recoil}) decreases with time, we then have an expanding
Robertson-Walker type Universe in the $(t,X)$ directions, and a static
universe in the transverse directions $(Y,Z)$:
\begin{equation}
\langle\langle ds^{2}\rangle\rangle=d\varsigma^{2}-a^{2}(\varsigma
)(dX)^{2}+\dots\label{rwrecoil}%
\end{equation}
where the \textquotedblleft cosmic time\textquotedblright\ $\varsigma$ is
defined as $d\varsigma^{2}=(1+2\sigma^{2}(t))dt^{2}$ and the $\dots$ denote
the static directions which will not be of relevance to us here. It is a
priori natural to consider an isotropic D-particle foam, in which the
D-particles recoil in random directions so that $\langle\langle u_{i}%
\rangle\rangle=0$, $\langle\langle u_{i}u_{j}\rangle\rangle=\sigma^{2}%
\delta_{ij}~.$ Asymptotically $a(\varsigma)$ monotonically tends to a constant
with $\varsigma$. In such a case one obtains an isotropic Robertson-Walker
like Universe. For simplicity in this work we concentrate on effective
two-dimensional universes, since such an assumption will not affect the
qualitative features of our approach. In this sense, the interaction of
D-particle foam with stringy matter on brane worlds, leads to an effective
space-time metric, \textquotedblleft felt\textquotedblright\ by the matter,
which, depending on properties of the foam, can have the form of homogeneous
and isotropic Robertson Walker cosmology.

\section{Effective Toy Model Field Theory of Bosons with Mixing
\label{sec:expansion}}

We will use the string theory considerations, leading to (\ref{rwrecoil}), as
a motivation for studying the possible consequences of our D-particle/string
hypothesis for flavour mixing within the context of a simple field theoretic
$1+1$ dimensional model with a horizon. Before starting, we mention that
flavour need not be conserved during the capture process of
fig.~\ref{fig:recoil}, in the sense that the re-emitted string, after the
capture, may have a different flavour (and mass) from the incident one.
However, the spatial momentum is conserved in the process. In this sense, our
D-particle foam may be considered as a medium of inducing flavour \emph{
mixing}~\cite{mavsark,barenboim}, which might be a feature of quantum gravity
foam$\varsigma$ in general.

As we shall discuss below, the effects of flavour changing on the space time
are drastic, and result in a ``new Fock-space vacuum'' state of the form
suggested in \cite{vitiello} for generic theories with mixing. As we shall
discuss in section \ref{sec:eos}, this results in non-trivial contributions to
the vacuum energy, of cosmological constant type, which come \emph{mostly}
from the infrared modes of the flavoured particles, of momenta less than the a
representative scale of order of the sum of the masses. These constant (in
time) contributions imply that, eventually, the flavour-changing process
\emph{dominate} the simple Robertson-Walker metric structure (\ref{metric2b}%
,{\ref{rwrecoil}) arising from the D-particle recoil distortions, results in
de Sitter type Universes. }

\subsection{Formalism in an Expanding Universe}

\bigskip\ Let us consider the metric in $1+1$ dimensions \cite{birrell}
\begin{equation}
ds^{2}=dt^{2}-a^{2}\left(  t\right)  dx^{2} \label{metric}%
\end{equation}
which in conformal time co-ordinates has the form we have
\begin{equation}
ds^{2}=C\left(  \eta\right)  \left(  d\eta^{2}-dx^{2}\right)  .
\label{metric2}%
\end{equation}
According to our discussion above, we first consider Robertson-Walker
expanding Universes, interpolating \emph{smoothly} between asymptotically flat
space times. In our D-particle-foam picture, leading to (\ref{metric2b}%
,\ref{rwrecoil}) this can be achieved by considering models of foam in which
the density of the D-particles in the bulk is not uniform but varies
appropriately in such a way that, as the D3-brane world moves through the
population of D-particles (c.f. fig.~\ref{fig:recoil}), the effective density
of the D-particles crossing the D3 brane varies smoothly with time so as to
lead to the above situation of smoothly expanding isotropic Universes.
Moreover, we assume for simplicity the case in which the fluctuation parameter
$\sigma^{2}(t) $ never reaches the horizon scale 1/2, but is always kept
sufficiently small. This is only for simplicity, since the essential features
of our expanding Universe, that would be relevant for our discussion in this
work, namely particle production, can already be captured in such horizon-free
situation. In any case, as we have already mentioned, and shall discus in
section \ref{sec:eos}, non-perturbative effects of our flavour vacuum will
produce de Sitter space times with asymptotic future horizons.

A simple choice for $C\left(  \eta\right)  $, capturing the above features is
\begin{equation}
C\left(  \eta\right)  =A+B\tanh\left(  \rho\eta\right)  \label{expansion}%
\end{equation}
with $A>B>0$ and $\rho>0$ and has the feature of $C\left(  \eta\right)  $
asymptotically tending to a constant as $\eta\rightarrow\infty$. This space
time is asymptotically accelerating, for sufficiently negative $\rho\eta
\to-\infty$, and decelerating for positive $\rho\eta\to\infty$; hence there
are no asymptotic future cosmic horizons. This is consistent with the absence
of such horizons in perturbative string theory~\cite{dine}, employed in our
world-sheet approach to D-particle foam, which is based on well-defined
Scattering-matrix elements and asymptotic states. In the D-particle picture
given above, if we assume that asymptotically in cosmic conformal time
$\eta\to\infty$, the recoil fluctuations $\sigma^{2} $ tend to a constant
value, $\widetilde{\sigma}^{2}$, we can identify $A\simeq1-4\widetilde{\sigma
}^{2}$ and $B=2\widetilde{\sigma}\,^{2}$. Of course, there is also a flat
space limit when is $B\rightarrow0$, i.e. vanishing D-particle recoil. The
space-time (\ref{expansion}) is considered as a \emph{perturbative} stringy
background on which we shall discuss flavour changing process resulting from
the stringy, topologically non-trivial processes of capture in our foam model,
depicted in fig.~\ref{fig:recoil}. As mentioned above, such flavour changing
process will lead to a new vacuum state, which in turn will imply
contributions to the vacuum energy of the Universe, of de Sitter type, that
will eventually dominate the space time (\ref{expansion}), leading to eternal
acceleration and asymptotic horizons. This should be considered as a
\emph{non-perturbative} contribution of the flavour-changing non-trivial
vacuum structure of our model.

On the other hand, if one would like to use an exactly soluble model, without
future horizons, interpolating between a de Sitter space time at a certain era
in the past and a flat Minkowski space time at the future asymptotic end of
cosmic time, then (s)he could consider as an instructive example the following
form of the scale factor~\cite{gubser}:
\begin{equation}
C(\eta) = A + B\frac{\eta}{\sqrt{\eta^{2} + \frac{1}{\rho^{2}}}}
\label{spacetime2}%
\end{equation}
with the model incorporating the de Sitter phase in the far past, $\rho\eta
\ll-1$ for $A=B >0, \rho> 0$. Indeed, for sufficiently negative conformal
times, such that $(\rho\eta)^{2} \gg1$, the model implies inflationary
evolution, since in that range of parameters
\begin{equation}
C(\eta) \simeq\frac{A}{2}\frac{1}{(\eta\rho)^{2}}~. \label{ds2}%
\end{equation}
On the other hand, in the asymptotic future $\eta\rho\gg1$, the model
asymptotes to flat Minkowski space time, with $C(\eta) \to2A$.

In what follows we shall not discuss explicitly the space time
(\ref{spacetime2}) further, but concentrate instead on (\ref{expansion}),
since for our qualitative purposes in this section of demonstrating the effect
of expansion insofar as particle production is concerned, the two space times
(\ref{spacetime2}) and (\ref{expansion}) are equivalent. However, we note
that, on anticipating a small Hubble parameter of our D-foam-induced-de Sitter
space time (c.f. (\ref{flavourenergy})), we are essentially interested in the
small $\rho$ regime in which both space-times (\ref{expansion}) and
(\ref{spacetime2}) undergo linear expansion with the (conformal) time. This
limit of slow expansion, $\rho\to0$ will be understood in what follows.

It goes without saying that the above issues are far from being considered as
resolved, especially in a string theory context, where the issues of an
inflationary phase, and in particular a smooth (``graceful'') exit from it,
are far from being understood analytically. In critical string theory, there
are recent attempts towards a construction of exact solutions, incorporating
inflation, by making use of the properties of the compact Calabi-Yau spaces in
concrete, phenomenologically semi-realistic (as far as incorporation of the
Standard model is concerned) brane world models~\cite{quevedo}. However, the
exit phase from the de Sitter era, together with attaining the very small
value for the vacuum energy observed today, are still not understood, in our
opinion. Some attempts towards these latter issues do exist, however, within
the context of non-critical (Liouville) string framework, which the present
D-particle foam model belongs to. For further details we refer the reader to
the literature~\cite{liouvinfl}.

In the present article, we shall certainly not attempt to offer any
quantitative resolution of these problems, given that our considerations below
will be based on effective field theories. Nevertheless, we think that our
findings are sufficiently interesting to probe further studies along this
direction, which might yield useful contributions to the quest for a proper
understanding of quantum gravity and its connection with other unresolved
issues in fundamental physics, such as the flavour mixing problem.

After this necessary explanatory remarks, we now proceed with our analysis. In
conformal co-ordinates our expanding-Universe background space time is written
as
\[
g_{\mu\nu}=\left(
\begin{array}
[c]{cc}%
C\left(  \eta\right)  & 0\\
0 & -C\left(  \eta\right)
\end{array}
\right)  ,\,\,g^{\mu\nu}=\left(
\begin{array}
[c]{cc}%
C^{-1}\left(  \eta\right)  & 0\\
0 & -C^{-1}\left(  \eta\right)
\end{array}
\right)
\]
with $\left\vert \det g_{\mu\nu}\right\vert \equiv g=C^{2}\left(  \eta\right)
$. We will consider the dynamics of a real scalar free field $\phi$
(appropriate for a neutral meson) of mass $m$ in this background:%
\begin{equation}
C^{-1}\left(  \eta\right)  \partial_{\mu}\left(  C\left(  \eta\right)
g^{\mu\nu}\partial_{\nu}\phi\right)  +m^{2}\phi=0 \label{KleinGordon}%
\end{equation}
This simplifies to
\begin{equation}
\left(  \partial_{\eta}^{2}-\partial_{x}^{2}\right)  \phi+m^{2}C\left(
\eta\right)  \phi=0. \label{KG1}%
\end{equation}
On writing \cite{birrell} $\phi\left(  \eta,x\right)  =U\left(  \eta\right)
V\left(  x\right)  $ we can find a complete basis set for expanding fields
from%
\begin{equation}
\frac{d^{2}}{dx^{2}}V+k^{2}V=0 \label{sepvar1}%
\end{equation}
and%
\begin{equation}
\frac{d^{2}}{d\eta^{2}}U+\left(  k^{2}+m^{2}C\left(  \eta\right)  \right)
U=0. \label{sepvar2}%
\end{equation}
We have $V\left(  x\right)  =e^{ikx}$ and $U(\eta)=c_{1}U_{1}(\eta)+c_{2}%
U_{2}(\eta)$ where $U_{1}(\eta)$ and $U_{2}(\eta)$ are given in terms of the
hypergeometric function $F\left(  a,b;c,z\right)  \left(  =\frac{\Gamma\left(
c\right)  }{\Gamma\left(  a\right)  \Gamma\left(  b\right)  }\sum
_{n=0}^{\infty}\frac{\Gamma\left(  a+n\right)  \Gamma\left(  b+n\right)
}{\Gamma\left(  c+n\right)  }\frac{z^{n}}{n!}\right)  $. It will be helpful to
introduce the notation
\begin{equation}
\omega_{in,k}=\sqrt{k^{2}+m^{2}\left(  A-B\right)  }, \label{freq1}%
\end{equation}%
\begin{equation}
\omega_{out,k}=\sqrt{k^{2}+m^{2}\left(  A+B\right)  }, \label{freq2}%
\end{equation}
and
\begin{equation}
\omega_{\pm,k}=\frac{1}{2}\left(  \omega_{out,k}\pm\omega_{in,k}\right)  .
\label{freq3}%
\end{equation}
In terms of these variables it can be shown that~\cite{birrell}%
\begin{align}
&  U_{1}\left(  \eta\right) \label{soln1}\\
&  \propto\left(  1+\tanh\left(  \rho\eta\right)  \right)  ^{\alpha_{1}%
-\frac{1}{2}}\left(  1-\tanh\left(  \rho\eta\right)  \right)  ^{\beta
_{1}-\frac{1}{2}}F\left(  \frac{i}{\rho}\omega_{+},1+\frac{i}{\rho}\omega
_{+};1+i\frac{\omega_{in}}{\rho},\frac{1}{2}\left(  1+\tanh\left(  \rho
\eta\right)  \right)  \right)
\end{align}
where $\alpha_{1}=\frac{1}{2}\left(  1+\frac{i\omega_{in}}{\rho}\right)
,\,\beta_{1}=\frac{1}{2}\left(  1+\frac{i\omega_{out}}{\rho}\right)  $. We can
find another independent solution $U_{2}\left(  \eta\right)  $ on making the
change $\eta\rightarrow-\eta$ and $B\rightarrow-B$ \ $\left(  \text{i.e.
}\omega_{in}\longleftrightarrow\omega_{out}\right)  $ and so
\begin{align}
&  U_{2}\left(  \eta\right) \label{soln2}\\
&  \propto\left(  1-\tanh\left(  \rho\eta\right)  \right)  ^{\alpha_{1}%
-\frac{1}{2}}\left(  1+\tanh\left(  \rho\eta\right)  \right)  ^{\beta
_{1}-\frac{1}{2}}F\left(  \frac{i}{\rho}\omega_{+},1+\frac{i}{\rho}\omega
_{+};1+i\frac{\omega_{out}}{\rho},\frac{1}{2}\left(  1-\tanh\left(  \rho
\eta\right)  \right)  \right)  .
\end{align}
We should note that complex conjugation of $U_{i}\left(  \eta\right)  $ is
also a solution of (\ref{sepvar2}) and this is effected by taking the complex
conjugations of the arguments of $F$ and also the exponents in (\ref{soln1})
and (\ref{soln2}). In terms of these solutions the appropriate \emph{in} and
\emph{out} solutions can be taken to be \
\begin{equation}
u_{k}^{in}\left(  \eta,x\right)  =e^{ikx}U_{1}^{\ast}\left(  \eta\right)
\label{instate}%
\end{equation}
and
\begin{equation}
u_{k}^{out}\left(  \eta,x\right)  =e^{ikx}U_{2}\left(  \eta\right)
\label{outstate}%
\end{equation}
where the constants of proportionality in (\ref{soln1}) and (\ref{soln2} are
chosen on the grounds of normalization. It is known that there are Bogolubov
coefficients $\alpha_{k}$ and $\beta_{k}$ such that~\cite{birrell}
\begin{equation}
u_{k}^{in}\left(  \eta,x\right)  =\alpha_{k}u_{k}^{out}\left(  \eta,x\right)
+\beta_{k}u_{-k}^{out\ast}\left(  \eta,x\right)  \label{Bcoeff1}%
\end{equation}
and
\begin{equation}
u_{k}^{out}\left(  \eta,x\right)  =\alpha_{k}^{\ast}u_{k}^{in}\left(
\eta,x\right)  +\beta_{k}u_{-k}^{in\ast}\left(  \eta,x\right)  .
\label{Bcoeff2}%
\end{equation}
with
\begin{equation}
\alpha_{k}=\sqrt{\frac{\omega_{out,k}}{\omega_{in,k}}}\frac{\Gamma\left(
1-\frac{i\omega_{in,k}}{\rho}\right)  \Gamma\left(  -\frac{i\omega_{out,k}%
}{\rho}\right)  }{\Gamma\left(  -\frac{i\omega_{+,k}}{\rho}\right)
\Gamma\left(  1-\frac{i\omega_{+,k}}{\rho}\right)  } \label{Bcoeff3}%
\end{equation}
and%
\begin{equation}
\beta_{k}=\sqrt{\frac{\omega_{out,k}}{\omega_{in,k}}}\frac{\Gamma\left(
1-\frac{i\omega_{in,k}}{\rho}\right)  \Gamma\left(  \frac{i\omega_{out,k}%
}{\rho}\right)  }{\Gamma\left(  \frac{i\omega_{-,k}}{\rho}\right)
\Gamma\left(  1+\frac{i\omega_{-,k}}{\rho}\right)  }. \label{Bcoeff4}%
\end{equation}
(N.B. the limit of no expansion is given by $B\rightarrow0$ with $\alpha
_{k}\rightarrow1$ and $\beta_{k}\rightarrow0$ as required.) This formalism
requires a slight generalization when there are $2$ distinct mass eigenstates.
We will need in the obvious way $\omega_{in}^{\left(  i\right)  }$,
$\omega_{out}^{\left(  i\right)  }$, $m^{\left(  i\right)  }$ with $i=1,2$
with $m$ replaced by $m^{\left(  i\right)  }$. In terms of creation $\left\{
a^{\dagger}\right\}  $ and annihilation $\left\{  a\right\}  $ operators for
massive particles in the \textit{in} and \textit{out} Hilbert spaces we have
the relations
\begin{align}
a_{k}^{in\,\left(  i\right)  }  &  =\alpha_{k}^{\left(  i\right)  \ast}%
a_{k}^{out\,\left(  i\right)  }-\beta_{k}^{\left(  i\right)  \ast}%
a_{-k}^{out\,\left(  i\right)  \dagger},\label{Bcoeff5}\\
a_{k}^{out\,\left(  i\right)  }  &  =\alpha_{k}^{\left(  i\right)  }%
a_{k}^{in\,\left(  i\right)  }+\beta_{k}^{\left(  i\right)  \ast}%
a_{-k}^{in\,\left(  i\right)  \dagger}. \label{Bcoeff6}%
\end{align}
where for $i=1,2$
\begin{equation}
\alpha_{k}^{\left(  i\right)  }=\sqrt{\frac{\omega_{out,k}^{\left(  i\right)
}}{\omega_{in,k}^{\left(  i\right)  }}}\frac{\Gamma\left(  1-\frac
{i\omega_{in,k}^{\left(  i\right)  }}{\rho}\right)  \Gamma\left(
-\frac{i\omega_{out,k}^{\left(  i\right)  }}{\rho}\right)  }{\Gamma\left(
-\frac{i\omega_{+,k}^{\left(  i\right)  }}{\rho}\right)  \Gamma\left(
1-\frac{i\omega_{+,k}^{\left(  i\right)  }}{\rho}\right)  }, \label{Bcoeff9}%
\end{equation}
and%
\begin{equation}
\beta_{k}^{\left(  i\right)  }=\sqrt{\frac{\omega_{out,k}^{\left(  i\right)
}}{\omega_{in,k}^{\left(  i\right)  }}}\frac{\Gamma\left(  1-\frac
{i\omega_{in,k}^{\left(  i\right)  }}{\rho}\right)  \Gamma\left(
\frac{i\omega_{out,k}^{\left(  i\right)  }}{\rho}\right)  }{\Gamma\left(
\frac{i\omega_{-,k}^{\left(  i\right)  }}{\rho}\right)  \Gamma\left(
1+\frac{i\omega_{-,k}^{\left(  i\right)  }}{\rho}\right)  }. \label{Bcoeff10}%
\end{equation}
The corresponding massive fields have the standard non-zero commutator%
\begin{equation}
\left[  \phi^{\left(  i\right)  }\left(  \eta,x\right)  ,\partial_{t\,}%
\phi^{\left(  j\right)  }\left(  \eta,y\right)  \right]  =i\delta\left(
x-y\right)  \delta_{ij}. \label{commutn}%
\end{equation}
It will be useful when discussing issues of orthogonality for condensate
states that we put our spatial system in a box of length $L$ with periodic
boundary conditions i.e. with $k=\frac{2\pi j}{L}$;$\ $in natural units
\begin{equation}
\phi^{\left(  i\right)  }\left(  \eta,x\right)  =\sum_{j=-\infty}^{\infty
}\frac{1}{\sqrt{2L\omega_{j}^{\left(  i\right)  }}}\left[  a_{j}^{\left(
i\right)  }\left(  \eta\right)  e^{-i\frac{2\pi j}{L}x}+a_{j}^{\left(
i\right)  \dagger}\left(  \eta\right)  e^{i\frac{2\pi j}{L}x}\right]
\label{field}%
\end{equation}
with $\omega_{j}^{\left(  i\right)  }=\sqrt{\left(  \frac{2\pi j}{L}\right)
^{2}+m^{\left(  i\right)  \,^{2}}}$and so
\begin{equation}
\left[  \phi^{\left(  i\right)  }\left(  \eta,x\right)  ,\partial_{t}%
\phi^{\left(  i\right)  }\left(  \eta,x^{\prime}\right)  \right]  =i\sum
_{j}\frac{1}{L}e^{i\frac{2\pi j}{L}\left(  x-x^{\prime}\right)  }\equiv
i\Delta_{L}\left(  x-x^{\prime}\right)  . \label{commutation}%
\end{equation}
We also introduce a short distance cut-off $\delta x$ (or lattice spacing) so
that we have a discrete lattice structure: $N\delta x=\frac{L}{2}$ where $N$
is an integer. There would then be a maximum $\left\vert j\right\vert
=N\left(  =\frac{L}{2\delta x}\right)  $ and so $-N\leq j\leq N$ . Similarly
with momentum $-\frac{\pi}{\delta x}$ and so $-N\leq j\leq N$. Space points
become discrete: $x\rightarrow n\delta x$ and $x^{\prime}\rightarrow
n^{\prime}\delta x$ so that $-N\leq n,n^{\prime}\leq N$. We can then define
$\Delta_{L,\delta x}\left(  x-x^{\prime}\right)  \equiv\frac{1}{L}\sum
_{j=-N}^{N}e^{i\left(  x-x^{\prime}\right)  \frac{2\pi j}{L}}$ as a lattice
analogue of $\Delta_{L}\left(  x-x^{\prime}\right)  $. Consequently on the
lattice
\begin{equation}
\phi^{\left(  i\right)  }\left(  n\delta x\right)  =\sum_{j=-N}^{N}\frac
{1}{\sqrt{2L\omega_{j}^{\left(  i\right)  }}}\left(  a_{j}^{\left(  i\right)
}e^{i\frac{\pi jn}{N}}e^{-i\omega_{j}^{\left(  i\right)  }\eta}+a_{j}^{\left(
i\right)  \dagger}e^{-i\frac{\pi jn}{N}}e^{i\omega_{j}^{\left(  i\right)
}\eta}\right)  \label{discrete}%
\end{equation}
with \ $-N\leq n\leq N$.

\subsection{Flavour mixing and Vacuum}

Although it would be desirable to rigorously demonstrate string matter flavour
changes due to D-particle capture this requires non-perturbative string field
theory or M theory. Since such a theory is not available we will consider a
field theoretic treatment where the mixing angle $\theta$ is put in by hand.
It is later when we consider the vacuum energy contribution due to D particles
that we will require a correlation between $\theta$ and D-particle recoil.
Following \cite{vitiello} flavour mixings are generated by~ $\mathcal{G}%
\left(  \theta\right)  $%
\begin{equation}
\mathcal{G}\left(  \theta\right)  =\exp\left(  i\theta\mathcal{S}\right)
\label{generator}%
\end{equation}
where $\mathcal{S}$ is given by \
\begin{equation}
\mathcal{S}=\frac{i}{2}\sum_{j=0}^{N}\mathfrak{S}_{j} \label{generator1}%
\end{equation}
where
\begin{align}
\mathfrak{S}_{j}  &  =\gamma_{j-}\left(  \widetilde{a}_{j}^{\left(  1\right)
}\widetilde{a}_{-j}^{\left(  2\right)  }+\widetilde{a}_{-j}^{\left(  1\right)
}\widetilde{a}_{j}^{\left(  2\right)  }\right)  -\gamma_{j-}\left(
\widetilde{a}_{j}^{\left(  1\right)  \dagger}\widetilde{a}_{-j}^{\left(
2\right)  \dagger}+\widetilde{a}_{-j,1}^{\left(  1\right)  \dagger}%
\widetilde{a}_{j}^{^{\left(  2\right)  }\dagger}\right) \label{generator2}\\
&  +\gamma_{j+}\left(  \widetilde{a}_{j}^{\left(  1\right)  }\widetilde{a}%
_{j}^{\left(  2\right)  \dagger}+\widetilde{a}_{-j}^{\left(  1\right)
}\widetilde{a}_{-j}^{\left(  2\right)  \dagger}\right)  -\gamma_{j+}\left(
\widetilde{a}_{j}^{\left(  1\right)  \dagger}\widetilde{a}_{j}^{\left(
2\right)  }+\widetilde{a}_{-j}^{\left(  1\right)  \dagger}\widetilde{a}%
_{-j}^{\left(  2\right)  }\right)  ,\nonumber
\end{align}
\ \ with%
\begin{equation}
\gamma_{j\pm}\equiv\sqrt{\frac{\omega_{j}^{\left(  1\right)  }}{\omega
_{j}^{\left(  2\right)  }}}\pm\sqrt{\frac{\omega_{j}^{\left(  2\right)  }%
}{\omega_{j}^{\left(  1\right)  }}} \label{gplusminus}%
\end{equation}
(and so $\gamma_{-j\pm}=\gamma_{j\pm}$) and $\widetilde{a}_{j}^{\left(
i\right)  }=a_{j}^{\left(  i\right)  }e^{-i\omega_{j}^{\left(  i\right)  }t}$.
There are of course additional superscripts associated with the \emph{in} and
\emph{out} Hilbert spaces. Clearly we have the standard commutation relations
$\left[  \widetilde{a}_{j}^{\left(  i\right)  },\widetilde{a}_{j^{\prime}%
}^{\dagger\left(  i^{\prime}\right)  }\right]  =\delta_{jj^{\prime}}%
\delta_{ii^{\prime}},$and $\left[  \widetilde{a}_{j}^{\left(  i\right)
},\widetilde{a}_{j^{\prime}}^{\left(  i^{\prime}\right)  }\right]  =\left[
\widetilde{a}_{j}^{\dag\left(  i\right)  },\widetilde{a}_{j^{\prime}}%
^{\dagger\left(  i^{\prime}\right)  }\right]  =0$ . Hence
\begin{equation}
\left[  \mathfrak{S}_{j},\mathfrak{S}_{j^{\prime}}\right]  =0,
\label{generator3}%
\end{equation}%
\begin{equation}
\mathcal{G}\left(  \theta\right)  =%
%TCIMACRO{\dprod \limits_{j=1}^{N}}%
%BeginExpansion
{\displaystyle\prod\limits_{j=1}^{N}}
%EndExpansion
\exp\left(  -\frac{\theta}{2}\mathfrak{S}_{j}\right)  , \label{generator4}%
\end{equation}
and
\begin{equation}
\mathcal{G}^{-1}\left(  \theta\right)  =%
%TCIMACRO{\dprod \limits_{j=1}^{N}}%
%BeginExpansion
{\displaystyle\prod\limits_{j=1}^{N}}
%EndExpansion
\exp\left(  \frac{\theta}{2}\mathfrak{S}_{j}\right)  . \label{generator5}%
\end{equation}
It is then straightforward to check that
\begin{equation}
\mathcal{G}^{-1}\left(  \theta\right)  \widetilde{a}_{\pm j}^{\left(
1\right)  }\mathcal{G}\left(  \theta\right)  =\cos\theta\,\widetilde{a}_{\pm
j}^{\left(  1\right)  }+\frac{1}{2}\sin\theta\,\left(  \gamma_{j-}%
\,\widetilde{a}_{\mp j}^{\dagger\left(  2\right)  }+\gamma_{j+\,}\widetilde
{a}_{\pm j}^{\left(  2\right)  }\right)  \label{flavour}%
\end{equation}
and so
\begin{equation}
\mathcal{G}^{-1}\left(  \theta\right)  a_{\pm j}^{\left(  1\right)
}\mathcal{G}\left(  \theta\right)  =\cos\theta\,a_{\pm j}^{\left(  1\right)
}+\frac{1}{2}\sin\theta\,\left(  \gamma_{j-}a_{\mp j}^{\dagger\left(
2\right)  }e^{i\left(  \omega_{j}^{\left(  1\right)  }+\omega_{-j}^{\left(
2\right)  }\right)  t}+\gamma_{j+}a_{\pm j}^{\left(  2\right)  }e^{i\left(
\omega_{j}^{\left(  1\right)  }-\omega_{j}^{\left(  2\right)  }\right)
t}\right)  . \label{flavour1}%
\end{equation}
Similarly we have (with $j>0$) :%
\begin{equation}
\mathcal{G}^{-1}\left(  \theta\right)  \widetilde{a}_{\pm j}^{\left(
2\right)  }\mathcal{G}\left(  \theta\right)  =\cos\theta\,\widetilde{a}_{\pm
j}^{\left(  2\right)  }+\frac{1}{2}\sin\theta\,\left(  \gamma_{j-\,}%
\widetilde{a}_{\mp j}^{\left(  1\right)  \dagger}-\gamma_{j+\,}\widetilde
{a}_{\pm j}^{\left(  1\right)  }\right)  . \label{flavour2}%
\end{equation}
These relations for creation and annihilation operators are consistent with
the following mixing relations for the corresponding massive fields
$\phi^{\left(  i\right)  }$:

\bigskip%

\begin{align}
\mathcal{G}^{-1}\left(  \theta\right)  \phi^{\left(  1\right)  }\left(
\eta,x\right)  \mathcal{G}\left(  \theta\right)   &  =\cos\theta
\,\,\phi^{\left(  1\right)  }\left(  \eta,x\right)  +\sin\theta\,\,\phi
^{\left(  2\right)  }\left(  \eta,x\right) \label{fieldflavour}\\
\mathcal{G}^{-1}\left(  \theta\right)  \phi^{\left(  2\right)  }\left(
\eta,x\right)  \mathcal{G}\left(  \theta\right)   &  =\cos\theta
\,\,\phi^{\left(  2\right)  }\left(  \eta,x\right)  -\sin\theta\,\,\phi
^{\left(  1\right)  }\left(  \eta,x\right)
\end{align}
the right hand sides of which can be identified with $\Phi_{\alpha}\left(
\eta,x\right)  $ and $\Phi_{\beta}\left(  \eta,x\right)  $ fields with flavour
quantum numbers $\alpha$ and $\beta$ \cite{vitiello}.The annihilation
operators $a_{j,\alpha}$ and $a_{j,\beta}$ for particles with flavour $\alpha$
and $\beta$ can be identified with $\mathcal{G}^{-1}\left(  \theta\right)
a_{j}^{\left(  1\right)  }\mathcal{G}\left(  \theta\right)  $ and
$\mathcal{G}^{-1}\left(  \theta\right)  a_{j}^{\left(  2\right)  }%
\mathcal{G}\left(  \theta\right)  $ respectively. Since we are interested in
the particular contributions of D-particles to mixing the above definition of
flavour vacua $\left\vert 0\right\rangle _{\alpha,\beta}$ will be restricted
to $k<k_{\max}=\frac{2\pi N^{\ast}}{L}$%
\begin{align}
a_{j,\alpha}\left\vert 0\right\rangle _{\alpha,\beta}  &  =0,\label{vac1}\\
a_{j,\beta}\left\vert 0\right\rangle _{\alpha,\beta}  &  =0 \label{vac2}%
\end{align}
on assuning that D-particle mechanism is responsible for particle mixing. For
values of $k$ higher than the cut-off $\left\vert 0\right\rangle
_{\alpha,\beta}$ will coincide with the massive vacuum i.e. $a_{\pm
j}^{\left(  i\right)  }\left\vert 0\right\rangle _{\alpha,\beta}=0$ for
$i=1,2$. This is an important difference between our model and that given in
\cite{vitiello}. Hence we can deduce that
\begin{equation}
\mathcal{G}_{\ast}\left(  \theta\right)  \left\vert 0\right\rangle
_{\alpha,\beta}=\left\vert 0\right\rangle _{1,2} \label{vac3}%
\end{equation}
where $\mathcal{G}_{\ast}\left(  \theta\right)  =%
%TCIMACRO{\dprod \limits_{j=1}^{N^{\ast}}}%
%BeginExpansion
{\displaystyle\prod\limits_{j=1}^{N^{\ast}}}
%EndExpansion
\exp\left(  -\frac{\theta}{2}\mathfrak{S}_{j}\right)  $ and so
\begin{equation}
\left\vert 0\right\rangle _{\alpha,\beta}=\mathcal{G}_{\ast}^{-1}\left(
\theta\right)  \left\vert 0\right\rangle _{1,2}. \label{vac4}%
\end{equation}
For finite $N^{\ast}$ the flavour and massive particle vacua have a non-zero
overlap. In the thermodynamic limit $N^{\ast}\rightarrow\infty$ this ceases to
be the case as will be discussed in the next section.

\bigskip

\subsection{\bigskip Relations between flavour and mass-eigenstate vacua}

We will be interested in the $N$ dependence of the overlap $f\left(
\theta\right)  $ between $\left\vert 0\right\rangle _{1,2}$ and $\left\vert
0\right\rangle _{\alpha,\beta}$ viz.%
\begin{equation}
f\left(  \theta\right)  =\left.  _{1,2}\right\langle 0\left\vert
\mathcal{G}_{\ast}^{-1}\left(  \theta\right)  \right.  \left\vert
0\right\rangle _{1,2} \label{overlap}%
\end{equation}
which has a factorized structure since $\mathcal{G}_{\ast}^{-1}\left(
\theta\right)  =%
%TCIMACRO{\dprod \limits_{j=1}^{N^{\ast}}}%
%BeginExpansion
{\displaystyle\prod\limits_{j=1}^{N^{\ast}}}
%EndExpansion
\mathcal{G}_{j}^{-1}\left(  \theta\right)  $ where $\mathcal{G}_{j}%
^{-1}\left(  \theta\right)  =\exp\left(  \frac{\theta}{2}\mathfrak{S}%
_{j}\right)  $. Hence it is sufficient to consider
\begin{equation}
f_{j}\left(  \theta\right)  =\left.  _{1,2}\right\langle 0\left\vert
\mathcal{G}_{j}^{-1}\left(  \theta\right)  \right.  \left\vert 0\right\rangle
_{1,2}. \label{overlap2}%
\end{equation}
Now
\begin{align}
\frac{d}{d\theta}f_{j}\left(  \theta\right)   &  =\frac{1}{2}\left.
_{1,2}\right\langle 0\left\vert \mathfrak{S}_{j}\mathcal{G}_{j}^{-1}\left(
\theta\right)  \right.  \left\vert 0\right\rangle _{1,2}\nonumber\\
&  =\frac{1}{2}\left.  _{1,2}\right\langle 0\left\vert \mathfrak{\gamma}%
_{j-}\left(  \widetilde{a}_{j}^{\left(  1\right)  }\widetilde{a}_{-j}^{\left(
2\right)  }+\widetilde{a}_{-j}^{\left(  1\right)  }\widetilde{a}_{j}^{\left(
2\right)  }\right)  \mathcal{G}_{j}^{-1}\left(  \theta\right)  \right.
\left\vert 0\right\rangle _{1,2} \label{overlap3}%
\end{align}
and%
\begin{align}
&  \left(  \widetilde{a}_{j}^{\left(  1\right)  }\widetilde{a}_{-j}^{\left(
2\right)  }+\widetilde{a}_{-j}^{\left(  1\right)  }\widetilde{a}_{j}^{\left(
2\right)  }\right)  \mathcal{G}_{j}^{-1}\left(  \theta\right) \\
&  =\mathcal{G}_{j}^{-1}\left(  \theta\right)  \left[  -\frac{1}{2}\sin
2\theta\,\,\mathfrak{\gamma}_{j-}+\frac{1}{4}\sin^{2}\theta
\,\,\mathfrak{\gamma}_{j-}^{2}\left(  \widetilde{a}_{j}^{\left(  1\right)
\dagger}\widetilde{a}_{-j}^{\left(  2\right)  \dagger}+\widetilde{a}%
_{-j}^{\left(  1\right)  \dagger}\widetilde{a}_{j}^{\left(  2\right)  \dag
}\right)  \right]  .\nonumber
\end{align}
Also, since $\left[  \mathfrak{S}_{j},\mathcal{G}_{j}^{-1}\left(
\theta\right)  \right]  =0$,
\begin{align}
\frac{d}{d\theta}f_{j}\left(  \theta\right)   &  =\frac{1}{2}\left.
_{1,2}\right\langle 0\left\vert \mathcal{G}_{j}^{-1}\left(  \theta\right)
\mathfrak{S}_{j}\right.  \left\vert 0\right\rangle _{1,2}\nonumber\\
&  =-\frac{1}{2}\mathfrak{\gamma}_{j-}\left.  _{1,2}\right\langle 0\left\vert
\mathcal{G}_{j}^{-1}\left(  \theta\right)  \left(  \widetilde{a}_{j}^{\left(
1\right)  \dagger}\widetilde{a}_{-j}^{\left(  2\right)  \dagger}+\widetilde
{a}_{-j}^{\left(  1\right)  \dagger}\widetilde{a}_{j}^{\left(  2\right)  \dag
}\right)  \right.  \left\vert 0\right\rangle _{1,2}. \label{overlap4}%
\end{align}
From \ref{overlap4} and \ref{overlap3} we deduce that
\begin{equation}
\left(  1+\frac{\mathfrak{\gamma}_{j-}^{2}}{4}\sin^{2}\theta\right)  \frac
{d}{d\theta}\,f_{j}\left(  \theta\right)  \,=-\frac{\mathfrak{\gamma}_{j-}%
^{2}}{4}\sin\left[  2\theta\right]  \,\,f_{j}\left(  \theta\right)  .
\label{overlap5}%
\end{equation}
which has a solution
\begin{equation}
\,f_{j}\left(  \theta\right)  \propto\left(  1+\frac{\mathfrak{\gamma}%
_{j-}^{2}}{4}\sin^{2}\left(  \theta\right)  \right)  ^{-1}.
\end{equation}
Because $\left(  m^{\left(  1\right)  2}-m^{\left(  2\right)  2}\right)  $ is
small, within the context of our model we can still choose a sufficiently
large $j$ $\left(  >N^{\prime}\text{ say}\right)  $ such that
\begin{equation}
\mathfrak{\gamma}_{j-}^{2}\simeq\frac{\left(  m^{\left(  1\right)
2}-m^{\left(  2\right)  2}\right)  L^{2}}{4\pi^{2}j^{2}}%
\end{equation}
and so%
\begin{equation}%
%TCIMACRO{\dprod \limits_{j=0}^{N^{\ast}}}%
%BeginExpansion
{\displaystyle\prod\limits_{j=0}^{N^{\ast}}}
%EndExpansion
\,f_{j}\left(  \theta\right)  \simeq%
%TCIMACRO{\dprod \limits_{j=N^{\prime}+1}^{N^{\ast}}}%
%BeginExpansion
{\displaystyle\prod\limits_{j=N^{\prime}+1}^{N^{\ast}}}
%EndExpansion
\frac{1}{1+\frac{\left(  m^{\left(  1\right)  2}-m^{\left(  2\right)
2}\right)  L^{2}}{4\pi^{2}j^{2}}\sin^{2}\theta}\times%
%TCIMACRO{\dprod \limits_{j=0}^{N^{\prime}}}%
%BeginExpansion
{\displaystyle\prod\limits_{j=0}^{N^{\prime}}}
%EndExpansion
\frac{1}{1+\frac{\mathfrak{\gamma}_{j-}^{2}}{4}\sin^{2}\theta}~.
\end{equation}
where $N^{\prime}\sim\frac{m^{\left(  1\right)  }m^{\left(  2\right)  }%
L}{\sqrt{2}\pi\left(  m^{\left(  1\right)  }+m^{\left(  2\right)  }\right)  }$.

\bigskip The factor $%
%TCIMACRO{\dprod \limits_{j=0}^{N^{\prime}}}%
%BeginExpansion
{\displaystyle\prod\limits_{j=0}^{N^{\prime}}}
%EndExpansion
\frac{1}{1+\frac{\mathfrak{\gamma}_{j-}^{2}}{4}\sin^{2}\theta}\equiv
F_{N^{\prime}}$ \ satisfies the bound%
\begin{equation}
F_{N^{\prime}}<%
%TCIMACRO{\dprod \limits_{j=1}^{N^{\prime}}}%
%BeginExpansion
{\displaystyle\prod\limits_{j=1}^{N^{\prime}}}
%EndExpansion
\frac{1}{1+a\left[  1-\frac{j^{2}}{N^{^{\prime}\,2}}\right]  }
\label{overlap6}%
\end{equation}
where $a\equiv\frac{1}{4}\sin^{2}\theta\,\frac{\left(  m^{\left(  1\right)
}-m^{\left(  2\right)  }\right)  ^{2}}{m^{\left(  1\right)  }m^{\left(
2\right)  }}$. The other factor in (\ref{overlap5}) can be written as
\[%
%TCIMACRO{\dprod \limits_{j=N^{\prime}+1}^{N^{\ast}}}%
%BeginExpansion
{\displaystyle\prod\limits_{j=N^{\prime}+1}^{N^{\ast}}}
%EndExpansion
\frac{1}{1+\frac{b}{j^{2}}}%
\]
where $b\equiv\frac{\left(  m^{\left(  1\right)  2}-m^{\left(  2\right)
2}\right)  L^{2}\sin^{2}\theta}{4\pi^{2}}$. Now we can define
\begin{equation}
\mathfrak{h}\left(  b,n\right)  =%
%TCIMACRO{\dprod \limits_{j=1}^{n}}%
%BeginExpansion
{\displaystyle\prod\limits_{j=1}^{n}}
%EndExpansion
\frac{1}{\left(  1+\frac{b}{j^{2}}\right)  } \label{factor}%
\end{equation}
and so
\begin{equation}%
%TCIMACRO{\dprod \limits_{j=N^{\prime}+1}^{N^{\ast}}}%
%BeginExpansion
{\displaystyle\prod\limits_{j=N^{\prime}+1}^{N^{\ast}}}
%EndExpansion
\frac{1}{1+\frac{b}{j^{2}}}=\frac{\mathfrak{h}\left(  b,N^{\ast}\right)
}{\mathfrak{h}\left(  b,N^{\prime}\right)  }.
\end{equation}
As $N^{\ast}\rightarrow\infty$ also $N^{\prime}\rightarrow\infty$ and so we
examine the asymptotic behaviour of $\mathfrak{h}\left(  b,n\right)  $. It can
be shown that
\begin{equation}
\mathfrak{h}\left(  b,n\right)  =\frac{\Gamma^{2}\left(  n+1\right)
\Gamma\left(  1-i\sqrt{b}\right)  \Gamma\left(  1+i\sqrt{b}\right)  }%
{\Gamma\left(  1-i\sqrt{b}+n\right)  \Gamma\left(  1+i\sqrt{b}+n\right)  }.
\end{equation}
From Stirling's formula as $n\rightarrow\infty$
\[
\frac{\Gamma^{2}\left(  n+1\right)  }{\Gamma\left(  1-i\sqrt{b}+n\right)
\Gamma\left(  1+i\sqrt{b}+n\right)  }\sim\left(  1-\frac{i\sqrt{b}}{n}\right)
^{-\frac{1}{2}+i\sqrt{b}}\left(  1+\frac{i\sqrt{b}}{n}\right)  ^{-\frac{1}%
{2}-i\sqrt{b}}\sim1.
\]
Thus $%
%TCIMACRO{\dprod \limits_{j=N^{\prime}+1}^{N^{\ast}}}%
%BeginExpansion
{\displaystyle\prod\limits_{j=N^{\prime}+1}^{N^{\ast}}}
%EndExpansion
\frac{1}{1+\frac{b}{j^{2}}}$ does not contribute to any orthogonality of the
flavour and mass vacua. An upper bound estimate $\mathfrak{F}_{N^{\prime
}\text{ }}$for $F_{N^{\prime}}$ given in (\ref{overlap6}), is
\begin{equation}
\mathfrak{F}_{N^{\prime}}=\left(  1+a\right)  ^{-N^{\prime}}%
%TCIMACRO{\dprod \limits_{j=1}^{N^{\prime}}}%
%BeginExpansion
{\displaystyle\prod\limits_{j=1}^{N^{\prime}}}
%EndExpansion
\frac{1}{1-\frac{a}{1+a}\frac{j^{2}}{N^{\prime\,\,2}}}. \label{overlap7}%
\end{equation}
Now it can be shown that
\begin{equation}
\log%
%TCIMACRO{\dprod \limits_{j=1}^{N^{\prime}}}%
%BeginExpansion
{\displaystyle\prod\limits_{j=1}^{N^{\prime}}}
%EndExpansion
\frac{1}{1-\frac{a}{1+a}\frac{j^{2}}{N^{\prime\,\,2}}}<N^{\prime}\left[
2+\log\left(  1+a\right)  -2\sqrt{\frac{a+1}{a}}\tanh^{-1}\left(  \sqrt
{\frac{a}{1+a}}\right)  \right]
\end{equation}
and since $a$ can reasonably be assumed to be very small and so
\begin{equation}
\log%
%TCIMACRO{\dprod \limits_{j=1}^{N^{\prime}}}%
%BeginExpansion
{\displaystyle\prod\limits_{j=1}^{N^{\prime}}}
%EndExpansion
\frac{1}{1-\frac{a}{1+a}\frac{j^{2}}{N^{\prime\,2}}}<N^{\prime}\left[
\frac{1}{3}a-\frac{7}{30}a^{2}+O\left(  a^{\frac{5}{2}}\right)  \right]  .
\end{equation}
From (\ref{overlap7})
\begin{equation}
\left\vert \mathfrak{F}_{N^{\prime}}\right\vert <\exp\left[  -\frac
{2a\,N^{\prime}}{3}\right]  \label{bound}%
\end{equation}
and so $\mathfrak{F}_{N^{\prime}}\rightarrow0$ as $N^{\prime}\rightarrow
\infty$. This demonstrates that the flavour and mass ground states are
orthogonal separately in the asymptotic \emph{in} and \emph{out} Hilbert
spaces provided $a$ is non-zero ( i.e. for non-zero $\sin^{2}\theta$ and
$\left(  m^{\left(  1\right)  }-m^{\left(  2\right)  }\right)  ^{2}$). This
orthogonality is actually robust to modifications of the energy-momentum
relation of the form $E^{2}=\frac{p^{2}+m^{2}}{1-\eta\left(  \frac{E}{M_{p}%
}\right)  ^{n}}$ due to possible quantum gravity effects where $n$ is a
positive integer \cite{dispersion}.

\section{\bigskip Vacua and the effect of expansion}

\subsection{Orthogonality}

The effect of expansion is well known to result in particle production for
co-moving observers. Consequently we shall consider its effect on the
orthogonality of the flavour and massive vacua. The relation between \emph{in}
\ and \emph{out }operators (c.f. (\ref{Bcoeff5}) and (\ref{Bcoeff6}))
connected by Bogoliubov coefficients can be effected in terms of an operator
$B$. In terms of%

\begin{equation}
\mathcal{S}\left(  v_{j}^{\left(  i\right)  }\right)  =\exp\left[
v_{j}^{\left(  i\right)  }a_{-j}^{\left(  i\right)  \dagger}a_{j}^{\left(
i\right)  \dagger}-v_{j}^{\left(  i\right)  \ast}a_{j}^{\left(  i\right)
}a_{-j}^{\left(  i\right)  }\right]  \label{squeezing}%
\end{equation}
and
\begin{equation}
P\left(  \phi_{j}^{\left(  i\right)  }\right)  =\exp\left(  -i\varphi
_{j}^{\left(  i\right)  }\left[  a_{j}^{\dagger\left(  i\right)  }%
a_{j}^{\left(  i\right)  }+a_{-j}^{\dagger\left(  i\right)  }a_{-j}^{\left(
i\right)  }\right]  \right)
\end{equation}
$B\left(  \varphi_{j}^{\left(  i\right)  },v_{j}^{\left(  i\right)  }\right)
$ can be written as $\mathcal{S}\left(  v_{j}^{\left(  i\right)  }\right)
P\left(  \varphi_{j}^{\left(  i\right)  }\right)  $. It can be shown that
\begin{align}
a_{j}^{\left(  i\right)  in}  &  =B^{\dagger}\left(  \varphi_{j}^{\left(
i\right)  out},v_{j}^{\left(  i\right)  out}\right)  a_{j}^{\left(  i\right)
out}B\left(  \varphi_{j}^{\left(  i\right)  out},v_{j}^{\left(  i\right)
out}\right) \label{bogolubov}\\
&  =\frac{v_{j}^{\left(  i\right)  out}e^{i\varphi_{j}^{\left(  i\right)
out}}}{\left\vert v_{j}^{\left(  i\right)  out}\right\vert }\sinh\left(
\left\vert v_{j}^{\left(  i\right)  out}\right\vert \right)  a_{-j}%
^{\dagger\left(  i\right)  out}+\cosh\left(  \left\vert v_{j}^{\left(
i\right)  out}\right\vert \right)  e^{-i\varphi_{j}^{\left(  i\right)  out}%
}a_{j}^{\left(  i\right)  out}. \label{bogolubov2}%
\end{align}
on identifying
\begin{align}
\frac{\widetilde{v}_{j}^{\left(  i\right)  out}}{\left\vert \widetilde{v}%
_{j}^{\left(  i\right)  out}\right\vert }\sinh\left(  \left\vert \widetilde
{v}_{j}^{\left(  i\right)  out}\right\vert \right)   &  =-\beta_{j}%
^{\ast\left(  i\right)  },\\
\cosh\left(  \left\vert \widetilde{v}_{j}^{\left(  i\right)  out}\right\vert
\right)  e^{-i\varphi_{j}^{\left(  i\right)  out}}  &  =\alpha_{j}^{\left(
i\right)  \ast}%
\end{align}
where $\widetilde{v}_{j}^{\left(  i\right)  out}\equiv v_{j}^{\left(
i\right)  out}e^{i\varphi_{j}^{\left(  i\right)  out}}$. Analogously
\begin{equation}
a_{j}^{\left(  i\right)  out}=\ \ \ B^{\dagger}\left(  \varphi_{j}^{\left(
i\right)  in},v_{j}^{\left(  i\right)  in}\right)  a_{j}^{\left(  i\right)
in}B\left(  \varphi_{j}^{\left(  i\right)  in},v_{j}^{\left(  i\right)
in}\right)
\end{equation}
with
\begin{equation}
\left\vert \widetilde{v}_{j}^{\left(  i\right)  in}\right\vert =\left\vert
\widetilde{v}_{j}^{\left(  i\right)  out}\right\vert =\tanh^{-1}\left[
\frac{\sinh\left(  \frac{\pi\omega_{-,j}^{\left(  i\right)  }}{\rho}\right)
}{\sinh\left(  \frac{\pi\omega_{+,j}^{\left(  i\right)  }}{\rho}\right)
}\right]
\end{equation}
where we have used the natural generalisations of (\ref{freq1},\ref{freq2} and
\ref{freq3}) defined by
\begin{equation}
\omega_{in,j}^{\left(  i\right)  }=\sqrt{\frac{4\pi^{2}j^{2}}{L^{2}%
}+m^{\left(  i\right)  2}\left(  A-B\right)  },
\end{equation}%
\begin{equation}
\omega_{out,j}^{\left(  i\right)  }=\sqrt{\frac{4\pi^{2}j^{2}}{L^{2}%
}+m^{\left(  i\right)  2}\left(  A+B\right)  },
\end{equation}
and
\begin{equation}
\omega_{\pm,j}^{\left(  i\right)  }=\frac{1}{2}\left(  \omega_{out,j}^{\left(
i\right)  }\pm\omega_{in,j}^{\left(  i\right)  }\right)  .
\end{equation}

\bigskip

Also
\begin{equation}
e^{i\varphi_{j}^{\left(  i\right)  out}}=\sqrt{\frac{\omega_{out,j}^{\left(
i\right)  }}{\omega_{in,j}^{\left(  i\right)  }}}\frac{\Gamma\left(
1+\frac{i\omega_{in,j}^{\left(  i\right)  }}{\rho}\right)  \Gamma\left(
\frac{i\omega_{out,j}^{\left(  i\right)  }}{\rho}\right)  }{\Gamma\left(
\frac{i\omega_{+,j}^{\left(  i\right)  }}{\rho}\right)  \Gamma\left(
1+\frac{i\omega_{+,j}^{\left(  i\right)  }}{\rho}\right)  \cosh\left(
\left\vert \widetilde{v}_{j}^{\left(  i\right)  in}\right\vert \right)  }.
\label{phase5}%
\end{equation}
Moreover $\widetilde{v}_{j}^{\left(  i\right)  in}=-\widetilde{v}_{j}^{\left(
i\right)  out}$ and $\varphi_{j}^{\left(  i\right)  in}=-\varphi_{j}^{\left(
i\right)  out}$ follows straightforwardly.

\bigskip

We have now the basic operators to calculate the overlap amplitude
\begin{equation}
\mathfrak{M}\left(  1,2;\alpha,\beta\right)  =\left.  _{1,2,in}\right\langle
\left.  0\right\vert \left.  0\right\rangle _{out,\alpha,\beta}
\label{newoverlap}%
\end{equation}
which is equivalent to
\begin{equation}
\mathfrak{M}\left(  1,2;\alpha,\beta\right)  =\left.  _{1,2,in}\right\langle
\left.  0\right\vert \mathcal{G}_{\ast}^{-1}\left(  \theta\right)  \left\vert
0\right\rangle _{out,1,2}. \label{newoverlap2}%
\end{equation}
The relation between \emph{in} and \emph{out} massive vacuum states
($\left\vert 0\right\rangle _{in1,2}$ and $\left\vert 0\right\rangle
_{out1,2}$) can be deduced from
\begin{equation}
a_{j}^{\left(  i\right)  in}\left\vert 0\right\rangle _{in1,2}=0=\ B^{\dagger
}\left(  \varphi_{j}^{\left(  i\right)  out},v_{j}^{\left(  i\right)
out}\right)  a_{j}^{\left(  i\right)  out}B\left(  \varphi_{j}^{\left(
i\right)  out},v_{k}^{\left(  i\right)  out}\right)  \left\vert 0\right\rangle
_{in,1,2}%
\end{equation}
since we require%
\begin{equation}
a_{j}^{\left(  i\right)  out}\left\vert 0\right\rangle _{out,1,2}=0\text{
\ }\forall\,j. \label{outvacuum}%
\end{equation}
Hence
\begin{equation}
\left\vert 0\right\rangle _{in,1,2}=%
%TCIMACRO{\dprod \limits_{i=1}^{2}}%
%BeginExpansion
{\displaystyle\prod\limits_{i=1}^{2}}
%EndExpansion%
%TCIMACRO{\dprod \limits_{j\geq0}}%
%BeginExpansion
{\displaystyle\prod\limits_{j\geq0}}
%EndExpansion
B^{\dagger}\left(  \varphi_{j}^{\left(  i\right)  out},v_{j}^{\left(
i\right)  out}\right)  \left\vert 0\right\rangle _{out,1,2}. \label{invac}%
\end{equation}
Similarly
\begin{equation}
\left\vert 0\right\rangle _{out,1,2}=%
%TCIMACRO{\dprod \limits_{i=1}^{2}}%
%BeginExpansion
{\displaystyle\prod\limits_{i=1}^{2}}
%EndExpansion%
%TCIMACRO{\dprod \limits_{j\geq0}}%
%BeginExpansion
{\displaystyle\prod\limits_{j\geq0}}
%EndExpansion
B^{\dagger}\left(  \varphi_{j}^{\left(  i\right)  in},v_{j}^{\left(  i\right)
in}\right)  \left\vert 0\right\rangle _{in,1,2}. \label{outvac}%
\end{equation}
$\therefore$
\[
\mathfrak{M}\left(  1,2;\alpha,\beta\right)  =\left.  _{1,2,out}\right\langle
\left.  0\right\vert
%TCIMACRO{\dprod \limits_{i=1}^{2}}%
%BeginExpansion
{\displaystyle\prod\limits_{i=1}^{2}}
%EndExpansion%
%TCIMACRO{\dprod \limits_{j\geq0}}%
%BeginExpansion
{\displaystyle\prod\limits_{j\geq0}}
%EndExpansion
B\left(  \varphi_{j}^{\left(  i\right)  out},v_{j}^{\left(  i\right)
out}\right)  \mathcal{G}_{_{\ast}out}^{-1}\left(  \theta\right)  \left\vert
0\right\rangle _{out,1,2}%
\]
where $\mathcal{G}_{_{\ast}out}^{-1}\left(  \theta\right)  \ $ is defined
analogously as before (c.f.. (\ref{generator})).
\begin{equation}
\mathcal{F}_{j}\left(  \theta\right)  =\left.  _{1,2,out}\right\langle \left.
0\right\vert \left\{
%TCIMACRO{\dprod \limits_{i=1}^{2}}%
%BeginExpansion
{\displaystyle\prod\limits_{i=1}^{2}}
%EndExpansion
B\left(  \varphi_{j}^{\left(  i\right)  out},v_{j}^{\left(  i\right)
out}\right)  \right\}  \mathcal{G}_{out,j}^{-1}\left(  \theta\right)
\left\vert 0\right\rangle _{out,1,2} \label{overlap1}%
\end{equation}
in terms of which
\begin{equation}
\mathfrak{M}\left(  1,2;\alpha,\beta\right)  =%
%TCIMACRO{\dprod \limits_{k>0}}%
%BeginExpansion
{\displaystyle\prod\limits_{k>0}}
%EndExpansion
\mathcal{F}_{k}\left(  \theta\right)  .
\end{equation}
In the small \textit{ }$\theta$ approximation ( appropriate for the size of
effects generated by quantum gravity)%
\begin{equation}
\mathcal{F}_{k}\left(  \theta\right)  \simeq\left.  _{1,2,out}\right\langle
\left.  0\right\vert \left\{
%TCIMACRO{\dprod \limits_{j=1}^{2}}%
%BeginExpansion
{\displaystyle\prod\limits_{j=1}^{2}}
%EndExpansion
\mathcal{S}\left(  v_{k}^{\left(  j\right)  out}\right)  \widetilde{P}\left(
\varphi_{k}^{\left(  j\right)  out}\right)  \right\}  \left(  1+\frac{\theta
}{2}\mathfrak{S}_{out,k}+\frac{\theta^{2}}{8}\mathfrak{S}_{out,k}^{2}\right)
\left\vert 0\right\rangle _{out,1,2}. \label{smalltheta}%
\end{equation}
In terms of the operators
\begin{align}
b_{+,j}^{\left(  i\right)  }  &  \equiv\frac{a_{-j}^{\left(  i\right)  }%
+a_{j}^{\left(  i\right)  }}{\sqrt{2}}\\
b_{-,j}^{\left(  i\right)  }  &  \equiv\frac{a_{-j}^{\left(  i\right)  }%
-a_{j}^{\left(  i\right)  }}{\sqrt{2}}%
\end{align}
(with $j>0$)
\begin{equation}
\mathcal{S}\left(  v_{j}^{\left(  i\right)  }\right)  =\exp\left[  \frac{1}%
{2}\left(  v_{j}^{\left(  i\right)  }b_{+,j}^{\left(  i\right)  \dagger
2}-v_{k}^{\left(  j\right)  \ast}b_{+,j}^{\left(  i\right)  2}\right)
\right]  \exp\left[  -\frac{1}{2}\left(  v_{j}^{\left(  i\right)  }%
b_{-,j}^{\left(  i\right)  \dagger2}-v_{j}^{\left(  i\right)  \ast}%
b_{-,j}^{\left(  i\right)  2}\right)  \right]  . \label{squeezeop1}%
\end{equation}
which has the form of a product of standard squeezing creation operators
$\mathbb{S}\left(  \xi\right)  =\exp\left[  \frac{1}{2}\left(  \xi
a^{\dagger2}-\xi^{\ast}a^{2}\right)  \right]  $ which generate a Fock space
representation of squeezed states
\begin{equation}
\mathbb{S}\left(  \xi\right)  \left\vert 0\right\rangle =\left(  1-\left\vert
\zeta\right\vert ^{2}\right)  ^{\frac{1}{4}}%
%TCIMACRO{\dsum \limits_{n=0}^{\infty}}%
%BeginExpansion
{\displaystyle\sum\limits_{n=0}^{\infty}}
%EndExpansion
\frac{\sqrt{\left(  2n\right)  !}}{2^{n}n!}\zeta^{n}\left\vert 2n\right\rangle
\label{squeezevac2}%
\end{equation}
where
\begin{equation}
\zeta=\frac{\xi}{\left\vert \xi\right\vert }\tanh\left(  \left\vert
\xi\right\vert \right)  .
\end{equation}
On using this property, in terms of $\zeta_{k}^{\left(  1\right)  out}%
\equiv\frac{v_{k}^{\left(  1\right)  out}}{\left\vert v_{k}^{\left(  1\right)
out}\right\vert }\tanh\left(  \left\vert v_{k}^{\left(  1\right)
out}\right\vert \right)  $ and $\zeta_{k}^{\left(  2\right)  out}\equiv
\frac{v_{k}^{\left(  2\right)  out}}{\left\vert v_{k}^{\left(  2\right)
out}\right\vert }\tanh\left(  \left\vert v_{k}^{\left(  2\right)
out}\right\vert \right)  $, it can be shown that
\begin{align}
&  \left.  _{1,2,out}\right\langle \left.  0\right\vert \left\{
%TCIMACRO{\dprod \limits_{j=1}^{2}}%
%BeginExpansion
{\displaystyle\prod\limits_{j=1}^{2}}
%EndExpansion
\mathcal{S}\left(  v_{k}^{\left(  j\right)  out}\right)  \widetilde{P}\left(
\phi_{k}^{\left(  j\right)  out}\right)  \right\}  \left\vert 0\right\rangle
_{out,1,2}\nonumber\\
&  =\sqrt{1-\left\vert \zeta_{k}^{\left(  1\right)  out}\right\vert ^{2}}%
\sqrt{1-\left\vert \zeta_{k}^{\left(  2\right)  out}\right\vert ^{2}}.
\label{thetasmallzero}%
\end{align}
Let us now compute the remaining terms in (\ref{smalltheta}). We can show
that
\begin{equation}
\left.  _{1,2,out}\right\langle \left.  0\right\vert \left\{
%TCIMACRO{\dprod \limits_{i=1}^{2}}%
%BeginExpansion
{\displaystyle\prod\limits_{i=1}^{2}}
%EndExpansion
\mathcal{S}\left(  v_{j}^{\left(  i\right)  out}\right)  P\left(  \phi
_{j}^{\left(  i\right)  out}\right)  \mathfrak{S}_{out,j}\right\}  \left\vert
0\right\rangle _{out,1,2}=0. \label{OrderOne}%
\end{equation}
The second order term in $\theta$ gives%
\begin{align}
&  \left.  _{1,2,out}\right\langle \left.  0\right\vert \left\{
%TCIMACRO{\dprod \limits_{i=1}^{2}}%
%BeginExpansion
{\displaystyle\prod\limits_{i=1}^{2}}
%EndExpansion
\mathcal{S}\left(  v_{j}^{\left(  i\right)  out}\right)  P\left(  \phi
_{j}^{\left(  i\right)  out}\right)  \mathfrak{S}_{out,j}^{2}\right\}
\left\vert 0\right\rangle _{out,1,2}\nonumber\\
&  =-2\gamma_{j,-}^{2}\sqrt{1-\left\vert \zeta_{j}^{\left(  1\right)
out}\right\vert ^{2}}\sqrt{1-\left\vert \zeta_{j}^{\left(  2\right)
out}\right\vert ^{2}}\left[  1-\frac{3}{4}e^{-2i\eta_{out,j}}\left(  \zeta
_{j}^{\left(  1\right)  out\ast}\zeta_{j}^{\left(  2\right)  out\ast}\right)
^{2}\right]  \label{SecondOrder}%
\end{align}
and so
\begin{equation}
\mathcal{F}_{j}\left(  \theta\right)  \simeq\sqrt{1-\left\vert \zeta
_{j}^{\left(  1\right)  out}\right\vert ^{2}}\sqrt{1-\left\vert \zeta
_{j}^{\left(  2\right)  out}\right\vert ^{2}}\left(  1-\frac{\theta^{2}}%
{4}\gamma_{j,-}^{2}\left[  1-\frac{3}{4}e^{-2i\eta_{out,j}}\left(  \zeta
_{j}^{\left(  1\right)  out\ast}\zeta_{j}^{\left(  2\right)  out\ast}\right)
^{2}\right]  \right)  . \label{smallthetaamplitude}%
\end{equation}
If the small $\theta$ expansion had not been made the total expression would
have been periodic in $\theta$ and so, to be compatible with the earlier
calculation ( in the absence of expansion), we can rewrite
(\ref{smallthetaamplitude}) as
\begin{equation}
\mathcal{F}_{j}\left(  \theta\right)  \simeq\frac{\sqrt{1-\left\vert \zeta
_{j}^{\left(  1\right)  out}\right\vert ^{2}}\sqrt{1-\left\vert \zeta
_{j}^{\left(  2\right)  out}\right\vert ^{2}}}{1+\gamma_{j,-}^{2}\frac
{\sin^{2}\theta}{4}\left(  1-\frac{3}{4}e^{-2i\eta_{out,j}}\left(  \zeta
_{j}^{\left(  1\right)  out\ast}\zeta_{j}^{\left(  2\right)  out\ast}\right)
^{2}\right)  }%
\end{equation}
which is again valid for small $\theta$. Now the denominator can be bounded
\begin{align*}
&  \left\vert 1+\gamma_{j,-}^{2}\frac{\sin^{2}\theta}{4}\left(  1-\frac{3}%
{4}e^{-2i\eta_{out,j}}\left(  \zeta_{j}^{\left(  1\right)  out\ast}\zeta
_{j}^{\left(  2\right)  out\ast}\right)  ^{2}\right)  \right\vert \\
&  >1+\gamma_{j,-}^{2}\frac{\sin^{2}\theta}{16}\left\{  4-3\left\vert
\zeta_{j}^{\left(  1\right)  out}\right\vert ^{2}\left\vert \zeta_{j}^{\left(
2\right)  out}\right\vert ^{2}\right\}  .
\end{align*}
Hence
%$\therefore$%
\begin{equation}
\left\vert \mathcal{F}_{j}\left(  \theta\right)  \right\vert ^{2}%
<\frac{\left(  1-\left\vert \zeta_{j}^{\left(  1\right)  out}\right\vert
^{2}\right)  \left(  1-\left\vert \zeta_{j}^{\left(  2\right)  out}\right\vert
^{2}\right)  }{\left(  1+\gamma_{j,-}^{2}\frac{\sin^{2}\theta}{16}\left\{
4-3\left\vert \zeta_{j}^{\left(  1\right)  out}\right\vert ^{2}\left\vert
\zeta_{j}^{\left(  2\right)  out}\right\vert ^{2}\right\}  \right)  ^{2}}.
\end{equation}
For large $j$,
\begin{equation}
\left\vert \zeta_{j}^{\left(  i\right)  out}\right\vert \simeq\frac{m^{\left(
i\right)  2}LB}{2\pi j\rho}e^{-\frac{2\pi^{2}j}{\rho L}}%
\end{equation}
which is a very sharp fall off with $j$. To leading order in $B$ for small $j$
and for small $\rho$ ( i.e. slow expansion)%
\begin{equation}
\left\vert \zeta_{j}^{\left(  i\right)  }\right\vert \simeq\exp\left(
-\frac{\pi}{2\rho A^{1/2}}\left[  \left(  2A-B\right)  m^{\left(  i\right)
}+\frac{4\pi^{2}j^{2}}{m^{\left(  i\right)  }L^{2}}\left(  1+\frac{B^{2}}%
{2A}\right)  \right]  \right)  .
\end{equation}
This behaviour for $\left\vert \zeta_{j}^{\left(  i\right)  }\right\vert $
does not change the previous arguments for orthogonality and so the flavour
vacuum (after expansion) and the original massive vacuum in the thermodynamic
limit are orthogonal. \ In \cite{vitiello} it was asserted that this
orthogonality implies that the two vacua are \emph{unitarily inequivalent},
i.e. there is no unitary transformation that connects them, and that the
Fock-space states constructed from one vacuum are orthogonal to the Fock-space
states constructed from the other. The lack of a unitary transformation
connecting the two vacua can be explicitly seen~\cite{ji} in our approach by
observing that the flavour vacuum is \emph{not} normalized to 1, as should be
the case if this vacuum were unitarily equivalent to the standard mass
eigenstate vacuum.

\subsection{Oscillation Probability}

Particle production during the expansion will lead to a modified oscillation
and we shall examine the signature of this modification within this model. In
the presence of expansion the transition amplitude $\mathfrak{M}_{osc}\left(
\mathcal{\alpha},k^{\prime};\mathcal{\beta},k\right)  $ for the oscillation of
flavour $\mathcal{\alpha}$ (momentum $k^{\prime}>0$) to $\mathcal{\beta}$
(momentum $k>0$) is
\begin{equation}
\mathfrak{M}_{osc}\left(  \mathcal{\alpha},k^{\prime};\mathcal{\beta}%
,k,\eta\right)  =\left.  _{k^{\prime},in}\right\langle \left.  \mathcal{\alpha
}\right\vert \left.  \mathcal{\beta}\left(  \eta\right)  \right\rangle
_{out,k}.
\end{equation}
\ \ \ where (by definition) $\left\vert \iota\left(  \eta\right)
\right\rangle _{out}=a_{\iota,\eta}^{\dagger out}\left\vert 0\right\rangle
_{out,\alpha,\beta}$ for $\iota=\alpha,\beta$ and similar definitions apply
for \emph{in }states. (Previous formulae with discrete momentum labels $j,$
when required, translate readily to the continuous $k$ label.) Hence, in terms
of the massive Fock space,
\begin{align}
\mathfrak{M}_{osc}\left(  \mathcal{\alpha},k^{\prime};\mathcal{\beta}%
,k,\eta\right)   &  =\left.  _{1,2,in}\right\langle \left.  0\right\vert
\left\{  \cos\left(  \theta\right)  \,a_{k^{\prime}}^{\left(  1\right)
in}+\frac{\sin\theta}{2}\left(  \gamma_{k^{\prime},-}a_{-k^{\prime}}%
^{\dagger\left(  2\right)  in}+\gamma_{k^{\prime},+}a_{k^{\prime}}^{\left(
2\right)  in}\right)  \right\} \\
&  \times\left\{  \cos\theta\;\widetilde{a}_{k}^{\left(  2\right)  out\dagger
}+\frac{\sin\theta}{2}\left(  \gamma_{k,-}\widetilde{a}_{-k}^{\left(
1\right)  out}-\gamma_{k,+}\widetilde{a}_{k}^{\left(  1\right)  out\dagger
}\right)  \right\}  \left\vert 0\right\rangle _{out,1,2}. \label{oscprob}%
\end{align}
It can be shown that
\begin{align}
&  \left\vert 0\right\rangle _{out,1,2}\nonumber\\
&  =%
%TCIMACRO{\dprod \limits_{\boldsymbol{k}\geq0}}%
%BeginExpansion
{\displaystyle\prod\limits_{\boldsymbol{k}\geq0}}
%EndExpansion
\left(
\begin{array}
[c]{c}%
\sqrt{1-\left\vert \zeta_{\boldsymbol{k}}^{\left(  1\right)  in}\right\vert
^{2}}\sqrt{1-\left\vert \zeta_{\boldsymbol{k}}^{\left(  2\right)
in}\right\vert ^{2}}\\
\times\sum_{n_{+,1}=0}^{\infty}\sum_{n_{+,2}=0}^{\infty}\sum_{n_{-,1}%
=0}^{\infty}\sum_{n_{+,1=0}}^{\infty}\mathfrak{g}_{\boldsymbol{k}}^{in}\left(
2n_{+,1},2n_{+,2},2n_{-,1},2n_{-,2}\right) \\
\left\vert 2n_{+,1},2n_{+,2},2n_{-,1},2n_{-,2}\right\rangle _{in}%
\end{array}
\right)  \label{outmassvac}%
\end{align}
where%
\begin{align}
&  \mathfrak{g}_{\boldsymbol{k}}^{in}\left(  2n_{+,1},2n_{+,2},2n_{-,1}%
,2n_{-,2}\right) \nonumber\\
&  \equiv\exp\left(  2i\left\{  \varphi_{\boldsymbol{k}}^{\left(  1\right)
in}\left[  n_{+,1}+n_{-,1}\right]  +\varphi_{\boldsymbol{k}}^{\left(
2\right)  in}\left[  n_{+,2}+n_{-,2}\right]  \right\}  \right) \nonumber\\
&  \times\zeta_{k}^{\left(  1\right)  in\,\left[  n_{+,1}+n_{-,1}\right]
}\zeta_{k}^{\left(  2\right)  in\,\left[  n_{+,2}+n_{-,2}\right]  }\frac
{\sqrt{\left(  2n_{+,1}\right)  !\left(  2n_{+,2}\right)  !\left(
2n_{-,1}\right)  !\left(  2n_{-,2}\right)  !}}{2^{n_{+,1}+n_{+,2}%
+n_{-,1}+n_{-,2}}\left(  n_{+,1}!\,n_{+,2}!\,n_{-,1}!\,n_{-,2}!\right)  }.
\end{align}
Hence
\begin{align}
&  \mathfrak{M}_{osc}\left(  \mathcal{\alpha},\boldsymbol{k}^{\prime
};\mathcal{\beta},\boldsymbol{k},\eta\right) \nonumber\\
&  =\left\{
%TCIMACRO{\dprod \limits_{k^{\prime\prime}\geq0}}%
%BeginExpansion
{\displaystyle\prod\limits_{k^{\prime\prime}\geq0}}
%EndExpansion
\sqrt{\left(  1-\left\vert \zeta_{k^{\prime\prime}}^{\left(  1\right)
}\right\vert ^{2}\right)  \left(  1-\left\vert \zeta_{k^{\prime\prime}%
}^{\left(  2\right)  }\right\vert ^{2}\right)  }\right\} \nonumber\\
&  \times\delta_{\boldsymbol{k}^{\prime}\boldsymbol{k}}\left\{  \varkappa
_{1,\boldsymbol{k}}\upsilon_{1,\boldsymbol{k}}\left(  \eta\right)
+\varkappa_{2,\boldsymbol{k}}\upsilon_{2,\boldsymbol{k}}\left(  \eta\right)
+\varkappa_{3,\boldsymbol{k}}\upsilon_{3,\boldsymbol{k}}\left(  \eta\right)
+\varkappa_{4,\boldsymbol{k}}\upsilon_{4,\boldsymbol{k}}\left(  \eta\right)
\right\}  \label{oscprobability}%
\end{align}
with $\varkappa_{1,\boldsymbol{k}}=\frac{1}{\sqrt{2}}\cos\theta=-\varkappa
_{2,\boldsymbol{k}}$, $\varkappa_{3,\boldsymbol{k}}=\frac{1}{2^{3/2}}%
\sin\theta\;\gamma_{\boldsymbol{k,+}}=-\varkappa_{4,\boldsymbol{k}}$ and
\[
\upsilon_{1,\boldsymbol{k}}\left(  \eta\right)  =\frac{1}{2^{3/2}}\sin
\theta\;\left(  \gamma_{\boldsymbol{k,-}}\beta_{-\boldsymbol{k}}^{\left(
1\right)  \ast}e^{-i\omega_{\boldsymbol{k}}^{\left(  1\right)  out}\eta
}-\gamma_{\boldsymbol{k,+}}\alpha_{\boldsymbol{k}}^{\left(  1\right)  \ast
}e^{i\omega_{\boldsymbol{k}}^{\left(  1\right)  out}\eta}\right)
=-\upsilon_{2,\boldsymbol{k}}\left(  \eta\right)
\]
and
\[
\upsilon_{3,\boldsymbol{k}}\left(  \eta\right)  =\frac{1}{\sqrt{2}}\cos
\theta\;\alpha_{\boldsymbol{k}}^{\left(  2\right)  \ast}e^{i\omega
_{\boldsymbol{k}}^{\left(  2\right)  out}\eta}=-\upsilon_{4,\boldsymbol{k}%
}\left(  \eta\right)  .
\]
(\ref{oscprobability}) simplifies to%
\begin{align}
&  \mathfrak{M}_{osc}\left(  \mathcal{\alpha},\boldsymbol{k}^{\prime
};\mathcal{\beta},\boldsymbol{k},\eta\right) \nonumber\\
&  =\frac{1}{4}\delta_{\boldsymbol{k}^{\prime}\boldsymbol{k}}\left\{
%TCIMACRO{\dprod \limits_{k^{\prime\prime}\geq0}}%
%BeginExpansion
{\displaystyle\prod\limits_{k^{\prime\prime}\geq0}}
%EndExpansion
\sqrt{\left(  1-\left\vert \zeta_{k^{\prime\prime}}^{\left(  1\right)
}\right\vert ^{2}\right)  \left(  1-\left\vert \zeta_{k^{\prime\prime}%
}^{\left(  2\right)  }\right\vert ^{2}\right)  }\right\} \nonumber\\
&  \times\sin2\theta\;\left[
\begin{array}
[c]{c}%
\gamma_{\boldsymbol{k},+}\left(  \alpha_{\boldsymbol{k}}^{\left(  2\right)
\ast}e^{i\omega_{\boldsymbol{k}}^{\left(  2\right)  out}\eta}-\alpha
_{\boldsymbol{k}}^{\left(  1\right)  \ast}e^{i\omega_{\boldsymbol{k}}^{\left(
1\right)  out}\eta}\right) \\
+\gamma_{\boldsymbol{k},-}\beta_{-\boldsymbol{k}}^{\left(  1\right)  \ast
}e^{-i\omega_{\boldsymbol{k}}^{\left(  1\right)  out}\eta}%
\end{array}
\right]  \label{oscprobability2}%
\end{align}
and can be readily interpreted as a modification of the usual formula for
oscillations with the extra momentum dependence (due to the field theoretic
treatment); now we have the \textit{additional} modification due to expansion
which is encoded in both the square root factor as well as the $\alpha$ and
$\beta$ coefficients. There is even a further effect in the oscillation
frequency due to the renormalization of the frequency represented by
$\omega^{out}$. We can readily estimate the values of $\alpha$ and $\beta$ for
both large and small $\left\vert k\right\vert $; for large $\left\vert
k\right\vert $ we find $\alpha_{k}\sim e^{-\frac{\pi\left\vert k\right\vert
}{2\rho}}\left(  \frac{\rho}{\left\vert k\right\vert }\right)  ^{\frac{1}{2}}$
and $\beta_{k}\sim-2i\pi^{2}\frac{m^{2}B}{\rho\left\vert k\right\vert
}e^{-\frac{\pi\left\vert k\right\vert }{\rho}}$ where the mass index has been suppressed.

\section{The equation of state for the flavour vacuum \label{sec:eos}}

The flavour vacuum as we have seen is orthogonal to the vacuum in terms of
mass eigenstates in the thermodynamic limit even in the presence of expansion.
Arguments based on d-particle scattering of stringy matter give support to the
flavour vacuum as the correct vacuum. By considering the expectation value of
the stress-energy tensor we can see to what extent it might play the role of
dark energy. This calculation, unlike previous ones, will consider a non-flat
metric which is of course necessary if there is an expanding universe. Our aim
is to investigate the theoretical possibility of a dark energy contribution
and the issues of D-particle capture and recoil, rather than phenomenological
relevance. Both early and late time expectation values are calculated.

The generic stress-energy tensor $T_{\mu\nu}$ for our scalar theory is
formally given by
\begin{equation}
T_{\mu\nu}=\frac{1}{2}\left(  \phi_{,\mu}\phi_{,\nu}+\phi_{,\nu}\phi_{,\mu
}\right)  -g_{\mu\nu}\frac{L}{\sqrt{-g}} \label{stressenergy}%
\end{equation}
where $L$ is the lagrangian density, $g=\det\left(  g_{\mu\nu}\right)  $ and
$\phi_{,\mu}=\frac{\partial\phi}{\partial x^{\mu}}$ ( the $x^{\mu}$ are the
generic space-time co-ordinates, $1+1$ dimensional in our case). There are of
course operator ordering ambiguities which we shall address later. In our
model
\begin{align}
T_{00}  &  =\sum_{i=1}^{2}\left\{  \frac{1}{2}\left(  \frac{\partial
\phi^{\left(  i\right)  }}{\partial\eta}\right)  ^{2}+\frac{1}{2}\left(
\frac{\partial\phi^{\left(  i\right)  }}{\partial x}\right)  ^{2}%
+\frac{m^{\left(  i\right)  2}C\left(  \eta\right)  }{2}\phi^{\left(
i\right)  2}\right\}  ,\\
T_{11}  &  =\sum_{i=1}^{2}\left\{  \frac{1}{2}\left(  \frac{\partial
\phi^{\left(  i\right)  }}{\partial x}\right)  ^{2}+\frac{1}{2}\left(
\frac{\partial\phi^{\left(  i\right)  }}{\partial\eta}\right)  ^{2}%
-\frac{m^{\left(  i\right)  2}C\left(  \eta\right)  }{2}\phi^{\left(
i\right)  2}\right\}  ,\\
T_{01}  &  =\frac{1}{2}\sum_{i=1}^{2}\left(  \frac{\partial\phi^{\left(
i\right)  }}{\partial\eta}\frac{\partial\phi^{\left(  i\right)  }}{\partial
x}+\frac{\partial\phi^{\left(  i\right)  }}{\partial x}\frac{\partial
\phi^{\left(  i\right)  }}{\partial\eta}\right)  .
\end{align}
We shall consider expectation values of $T_{\mu\nu}$ for the asymptotic early
and late times. For reasons of rigour we will work in the discrete momentum
basis. In this basis (suppressing $in$ and $out$ indices) let us define
\begin{equation}
b_{j}^{\left(  i\right)  }\left(  \eta,x\right)  \equiv-a_{j}^{\left(
i\right)  }e^{i\left(  \frac{2\pi j}{L}x-\omega_{j}^{\left(  i\right)  }%
\eta\right)  }+a_{j}^{\left(  i\right)  \dagger}e^{-i\left(  \frac{2\pi j}%
{L}x-\omega_{j}^{\left(  i\right)  }\eta\right)  }%
\end{equation}
and%
\begin{equation}
d_{j}^{\left(  i\right)  }\left(  \eta,x\right)  \equiv a_{j}^{\left(
i\right)  }e^{i\left(  \frac{2\pi j}{L}x-\omega_{j}^{\left(  i\right)  }%
\eta\right)  }+a_{j}^{\left(  i\right)  \dagger}e^{-i\left(  \frac{2\pi j}%
{L}x-\omega_{j}^{\left(  i\right)  }\eta\right)  }.
\end{equation}
In terms of these operators
\begin{align}
T_{00}  &  =\frac{1}{2}T_{00}^{\left(  1\right)  }+\frac{1}{2}T_{00}^{\left(
2\right)  }+\frac{C\left(  \eta\right)  }{2}T_{00}^{\left(  3\right)
},\label{stresseng1}\\
T_{11}  &  =\frac{1}{2}T_{00}^{\left(  2\right)  }+\frac{1}{2}T_{00}^{\left(
1\right)  }-\frac{C\left(  \eta\right)  }{2}T_{00}^{\left(  3\right)
},\label{stresseng2}\\
T_{01}  &  =\pi\sum_{i=1}^{2}\sum_{j,\,j^{\prime}}\nu_{jj^{\prime}}^{\left(
i\right)  }b_{j}^{\left(  i\right)  }\left(  \eta,x\right)  b_{j^{\prime}%
}^{\left(  i\right)  }\left(  \eta,x\right)  \label{stresseng3}%
\end{align}
where%
\begin{align}
T_{00}^{\left(  1\right)  }  &  =-\sum_{i=1}^{2}\left(  \sum_{j}\sqrt
{\frac{\omega_{j}^{\left(  i\right)  }}{2L}}b_{j}^{\left(  i\right)  }\left(
\eta,x\right)  \right)  ^{2},\label{stresseng4}\\
T_{00}^{\left(  2\right)  }  &  =-4\pi^{2}\sum_{i=1}^{2}\left(  \sum_{j}%
\frac{j}{\sqrt{2L^{3}\omega_{j}^{\left(  i\right)  }}}b_{j}^{\left(  i\right)
}\left(  \eta,x\right)  \right)  ^{2},\label{stresseng5}\\
T_{00}^{\left(  3\right)  }  &  =-\sum_{i=1}^{2}\left(  \sum_{j}%
\frac{m^{\left(  i\right)  }}{\sqrt{2L\omega_{j}^{\left(  i\right)  }}}%
d_{j}^{\left(  i\right)  }\left(  \eta,x\right)  \right)  ^{2},
\label{stresseng6}%
\end{align}
and $\nu_{jj^{\prime}}^{\left(  i\right)  }=\frac{1}{2L^{2}}\left(
\sqrt{\frac{\omega_{j}^{\left(  i\right)  }}{\omega_{j^{\prime}}^{\left(
i\right)  }}}j^{\prime}+\sqrt{\frac{\omega_{j}^{\left(  i\right)  }}%
{\omega_{j^{\prime}}^{\left(  i\right)  }}}j^{\prime}\right)  $.

We will now calculate $\left.  _{in,\alpha,\beta}\right\langle 0\left\vert
T_{\mu\nu}^{out}\left\vert 0\right\rangle _{\alpha,\beta,in}\right.  $. It is
straightforward to show that
\begin{equation}
\left.  _{in,\alpha,\beta}\right\langle 0\left\vert b_{j}^{\left(  i\right)
out}\left(  \eta,x\right)  b_{j^{\prime}}^{\left(  i\right)  out}\left(
\eta,x\right)  \left\vert 0\right\rangle _{\alpha,\beta,in}\right.
=\delta_{jj^{\prime}}\mathfrak{f}_{j,i}^{\left(  +\right)  }+\delta
_{j\,\,-j^{\prime}}\mathfrak{f}_{j,i}^{\left(  -\right)  } \label{matrixel1}%
\end{equation}
and
\begin{equation}
\left.  _{in,\alpha,\beta}\right\langle 0\left\vert d_{j}^{\left(  i\right)
out}\left(  \eta,x\right)  d_{j^{\prime}}^{\left(  i\right)  out}\left(
\eta,x\right)  \left\vert 0\right\rangle _{\alpha,\beta,in}\right.
=-\delta_{jj^{\prime}}\mathfrak{f}_{j,i}^{\left(  +\right)  }+\delta
_{j\,\,-j^{\prime}}\mathfrak{f}_{j,i}^{\left(  -\right)  }. \label{matrixel2}%
\end{equation}
The expressions for $\mathfrak{f}_{j,i}^{\left(  \pm\right)  }$ are given in
the Appendix . There are infinities in the expectation value of the stress
tensor which have conventionally been removed by renormalization. In flat
space-time this has been achieved by a suitable normal ordering. Certainly if
we consider our model conventionally as a quantum field theory in a time
dependent metric then we would have to renormalise~\cite{birrell}. This would
involve state dependent counter terms which, in a covariant procedure, can be
tensorially constructed from the metric tensor. However our model, as we have
seen, is motivated from D-particle capture of stringy matter and so its
interpretation cannot be completely that of a conventional field theory. We
will look at both the low and high momenta behaviour of $\mathfrak{f}%
_{j,i}^{\left(  \pm\right)  }$. The high momenta behaviour will indicate the
necessity of `conventional' renormalization since the flavour vacuum is
indistinguishable from the standard massive vacuum the while the low momenta
behaviour will give the contribution adapted to the D-particle capture approach.

\bigskip

\ The results can be summarised as:%
\begin{equation}
\left.  _{in,\alpha,\beta}\right\langle 0\left\vert T_{00}^{\left(  1\right)
out}\left\vert 0\right\rangle _{\alpha,\beta,in}\right.  =-\frac{1}%
{2L}\left\{  \sum_{j}\sum_{i=1}^{2}\omega_{j}^{\left(  i\right)  }\left(
\mathfrak{f}_{j,i}^{\left(  +\right)  }+\mathfrak{f}_{j,i}^{\left(  -\right)
}\right)  \right\}  , \label{contribn1}%
\end{equation}

\begin{equation}
\left.  _{in,\alpha,\beta}\right\langle 0\left\vert T_{00}^{\left(  2\right)
out}\left\vert 0\right\rangle _{\alpha,\beta,in}\right.  =-4\pi^{2}\sum
_{j}\sum_{i=1}^{2}\frac{j^{2}}{2L^{3}\omega_{j}^{\left(  i\right)  }}\left\{
\mathfrak{f}_{j,i}^{\left(  +\right)  }-\mathfrak{f}_{j,i}^{\left(  -\right)
}\right\}  , \label{contribn2}%
\end{equation}

\begin{equation}
\left.  _{in,\alpha,\beta}\right\langle 0\left\vert T_{00}^{\left(  3\right)
out}\left\vert 0\right\rangle _{\alpha,\beta,in}\right.  =\frac{1}{2L}\sum
_{j}\sum_{i=1}^{2}\frac{m^{\left(  i\right)  2}}{\omega_{j}^{\left(  i\right)
}}\left(  -\mathfrak{f}_{j,i}^{\left(  -\right)  }+\mathfrak{f}_{j,i}^{\left(
+\right)  }\right)  , \label{contribn3}%
\end{equation}
and%
\begin{equation}
\left.  _{in,\alpha,\beta}\right\langle 0\left\vert T_{01}^{out}\left\vert
0\right\rangle _{\alpha,\beta,in}\right.  =0. \label{contribn4}%
\end{equation}
Hence%
\begin{align}
\left.  _{in,\alpha,\beta}\right\langle 0\left\vert T_{00}^{out}\left\vert
0\right\rangle _{\alpha,\beta,in}\right.   &  =\frac{1}{2}\left.
_{in,\alpha,\beta}\right\langle 0\left\vert T_{00}^{\left(  1\right)
out}\left\vert 0\right\rangle _{\alpha,\beta,in}\right.  +\frac{1}{2}\left.
_{in,\alpha,\beta}\right\langle 0\left\vert T_{00}^{\left(  2\right)
out}\left\vert 0\right\rangle _{\alpha,\beta,in}\right. \nonumber\\
&  +\frac{C\left(  \eta\right)  }{2}\left.  _{in,\alpha,\beta}\right\langle
0\left\vert T_{00}^{\left(  3\right)  out}\left\vert 0\right\rangle
_{\alpha,\beta,in}\right. \nonumber\\
&  =-\frac{1}{4L}\sum_{j}\sum_{i=1}^{2}\omega_{j}^{\left(  i\right)  }\left(
\mathfrak{f}_{j,i}^{\left(  +\right)  }+\mathfrak{f}_{j,i}^{\left(  -\right)
}\right) \nonumber\\
&  -2\pi^{2}\sum_{j}\sum_{i=1}^{2}\frac{j^{2}}{2L^{3}\omega_{j}^{\left(
i\right)  }}\left\{  \mathfrak{f}_{j,i}^{\left(  +\right)  }-\mathfrak{f}%
_{j,i}^{\left(  -\right)  }\right\} \nonumber\\
&  +\frac{C\left(  \eta\right)  }{4L}\sum_{j}\sum_{i=1}^{2}\frac{m^{\left(
i\right)  2}}{\omega_{j}^{\left(  i\right)  }}\left(  \mathfrak{f}%
_{j,i}^{\left(  +\right)  }-\mathfrak{f}_{j,i}^{\left(  -\right)  }\right)
\label{energy}%
\end{align}
and
\begin{align}
\left.  _{in,\alpha,\beta}\right\langle 0\left\vert T_{11}^{out}\left\vert
0\right\rangle _{\alpha,\beta,in}\right.   &  =\frac{1}{2}\left.
_{in,\alpha,\beta}\right\langle 0\left\vert T_{00}^{\left(  2\right)
\,out}\left\vert 0\right\rangle _{\alpha,\beta,in}\right.  +\frac{1}{2}\left.
_{in,\alpha,\beta}\right\langle 0\left\vert T_{00}^{\left(  1\right)
\,out}\left\vert 0\right\rangle _{\alpha,\beta,in}\right. \nonumber\\
&  -\frac{C\left(  \eta\right)  }{2}\left.  _{in,\alpha,\beta}\right\langle
0\left\vert T_{00}^{(3)\,out}\left\vert 0\right\rangle _{\alpha,\beta
,in}\right. \nonumber\\
&  =-2\pi^{2}\sum_{j}\sum_{i=1}^{2}\frac{j^{2}}{2L^{3}\omega_{j}^{\left(
i\right)  }}\left\{  \mathfrak{f}_{j,i}^{\left(  +\right)  }-\mathfrak{f}%
_{j,i}^{\left(  -\right)  }\right\} \nonumber\\
&  -\frac{1}{4L}\left\{  \sum_{j}\sum_{i=1}^{2}\omega_{j}^{\left(  i\right)
}\left(  \mathfrak{f}_{j,i}^{\left(  +\right)  }+\mathfrak{f}_{j,i}^{\left(
-\right)  }\right)  \right\} \nonumber\\
&  -\frac{C\left(  \eta\right)  }{4L}\sum_{j}\sum_{i=1}^{2}\frac{m^{\left(
i\right)  2}}{\omega_{j}^{\left(  i\right)  }}\left(  -\mathfrak{f}%
_{j,i}^{\left(  -\right)  }+\mathfrak{f}_{j,i}^{\left(  +\right)  }\right)  .
\label{pressure}%
\end{align}

\bigskip The expressions in (\ref{energy}) and (\ref{pressure}) can be normal
ordered with respect to a suitable vacuum. Unlike elementary field theory
there is more than one obvious choice. The equation of state will vary
depending on this choice and can give physical insight into the fluid like
properties of the unusual flavour vacuum in the presence of an expanding
universe; so a normal ordered expectation can be defined as
\begin{equation}
\left.  _{in,\alpha,\beta}\right\langle 0\left\vert \colon T_{11}^{out}%
\colon\left\vert 0\right\rangle _{\alpha,\beta,in}\right.  =\left.
_{in,\alpha,\beta}\right\langle 0\left\vert T_{11}^{out}\left\vert
0\right\rangle _{\alpha,\beta,in}\right.  -\left\langle \Psi\right\vert
T_{11}^{out}\left\vert \Psi\right\rangle \label{normalorder}%
\end{equation}
where $\left\vert \Psi\right\rangle $ is a suitable `vacuum' state.We should
remark that various types of `normal'$\,$orderings can be accommodated through
the choice of $\left\vert \Psi\right\rangle $. The usual one in Minkowski
space-time reads:
\begin{align}
\colon b_{j}^{\left(  i\right)  }\left(  \eta,x\right)  b_{j^{\prime}%
}^{\left(  i\right)  }\left(  \eta,x\right)  \colon &  =b_{j}^{\left(
i\right)  }\left(  \eta,x\right)  b_{j^{\prime}}^{\left(  i\right)  }\left(
\eta,x\right)  +\delta_{jj^{\prime}},\label{normal1}\\
\colon d_{j}^{\left(  i\right)  }\left(  \eta,x\right)  d_{j^{\prime}%
}^{\left(  i\right)  }\left(  \eta,x\right)  \colon &  =d_{j}^{\left(
i\right)  }\left(  \eta,x\right)  d_{j^{\prime}}^{\left(  i\right)  }\left(
\eta,x\right)  -\delta_{jj^{\prime}}. \label{normal2}%
\end{align}
In our D-particle-foam model, which motivates our use of the \textquotedblleft
flavour\textquotedblright\ vacuum, the physically correct normal ordering is
dictated by the underlying microscopic physics of the foamy space-time at
Planck scales, and the non-trivial interactions with it of \emph{some} of the
flavour momentum modes, specifically the low-energy ones. A cloud of
D-particles induces a metric which is proportional to $\sigma^{2}$ (c.f.
(\ref{metric2b}), (\ref{rwrecoil})). In our picture, the vacuum contribution
to the expectation value of the stress tensor is due to the D-particle
recoil-velocity (statistical) Gaussian fluctuations (\ref{gaussian}), i.e. we
subtract away the $\sigma$ independent contribution, following our earlier
treatment of the induced stochastic light-cone fluctuations due to the
recoil-velocity fluctuations~\cite{recoil2}. This is consistent with the
recovery of the Minkowski, Lorentz-invariant situation (c.f. (\ref{metric2b}))
in the absence of D-particle recoil-velocity effects, i.e. the vanishing of
the vacuum expectation value of the matter stress-energy tensor in the limit
$\sigma^{2} \to0$.

We should recall here that for a relativistic perfect fluid%
\[
T_{\mu\nu}=-pg_{\mu\nu}+\left(  p+\rho\right)  v_{\mu}v_{\nu}%
\]
where $v_{\mu}$ is the $4$-velocity of the observer. From (\ref{contribn1}%
),(\ref{contribn2}), (\ref{contribn3}) and (\ref{stresseng1}),
(\ref{stresseng2}), (\ref{stresseng3}) we have the form of the expectation
values of the various components of the stress-energy tensor, and from those
we can conclude about the equation of state of our boson gas and check whether
$p=w\rho$ with $w=-1$, the expected form for cosmological constant. In order
to obtain finite expressions for $p$ and $\rho$ we would need some form of
momentum cut-off for the low $k$ modes. However, on identifying (in a
co-moving cosmological frame) $p=\left.  _{in,\alpha,\beta}\right\langle
0\left\vert \colon T_{11}^{out}\colon\left\vert 0\right\rangle _{\alpha
,\beta,in}\right.  $ and $\rho=\left.  _{in,\alpha,\beta}\right\langle
0\left\vert \colon T_{00}^{out}\colon\left\vert 0\right\rangle _{\alpha
,\beta,in}\right.  $ and on adopting the $\sigma$ independent subtraction
procedure, we find interestingly (c.f.. (\ref{energy}) and (\ref{pressure}))
that
\begin{equation}
\label{eos2}w=-1~.
\end{equation}
It is important to notice that (\ref{eos2}) has been obtained by a subtraction
procedure which was adopted \emph{independently} of the particular choice of
the expanding universe conformal factor $C(\eta)$. This is important, in that
it allows the back reaction effects of the flavour vacuum onto the background
space time to be incorporated self-consistently in our approach.

The determination of $p$ and $\rho$, which are vacuum expectation values,
involves a tacit averaging over the time scale $\tau$ in our vacuum. We
earlier estimated $\tau$ to be extremely short and so a time average over it
is equivalent to putting $\eta=0$ in the expressions. The contribution to the
cosmological constant from this vacuum is given by $\rho$.

It must be noted at this staged that the above calculations have been
performed in the regime in which the expansion rate of the Universe is slow,
and the space-time is almost a Minkowskian one. Hence the
cosmological-constant like equation of state that characterises our
Lorentz-violating vacuum pertains only to late eras of the Universe. In
general, one might encounter situations with a time dependent equation of
state and a time varying dark energy (i.e. quintessence).

We have earlier argued that the capture process is likely to be most efficient
for stringy matter with small $k$. In order to get an estimate for the
momentum cut-off of the capture process, and thus of the flavoured-vacuum
contribution to the cosmological constant, we can adopt an argument in
\cite{boyanovsky} by considering the scale inherent in the $k$ \ dependence of
the single particle momentum distributions in the flavour vacuum
\begin{equation}
n_{\iota}\left(  k,\eta\right)  \equiv\left.  _{\alpha,\beta}\right\langle
0\left\vert \widetilde{a}_{\iota,k}^{\dag}\left(  \eta,x\right)  \widetilde
{a}_{\iota,k}\left(  \eta,x\right)  \left\vert 0\right\rangle _{\alpha,\beta
}\right.  \label{singleparticle}%
\end{equation}
for $\iota=\alpha,\beta$. It can be shown that \cite{ji} $\,$%
\begin{equation}
n_{\alpha,k}\left(  t\right)  =\gamma_{k,-}^{2}\left[  {\frac{{\sin^{4}\theta
}}{4}\,\gamma_{k,+}^{2}\sin^{2}\omega_{in,k}^{\left(  2\right)  }\eta
+\frac{{\sin^{2}2\theta}}{4}\sin^{2}\left(  {\omega_{in,k}^{\left(  1\right)
}+\omega_{in,k}^{\left(  2\right)  }}\right)  \eta}\right]
\label{singleparticle2}%
\end{equation}
and%

\begin{equation}
n_{\beta,k}\left(  t\right)  =\gamma_{k,-}^{2}\left[  {\frac{{\sin^{4}\theta}%
}{4}\,\gamma_{k,+}^{2}\sin^{2}\omega_{in,k}^{\left(  1\right)  }\eta
+\frac{{\sin^{2}2\theta}}{4}\sin^{2}\left(  {\omega_{in,k}^{\left(  1\right)
}+\omega_{in,k}^{\left(  2\right)  }}\right)  \eta}\right]  .
\label{singleparticle3}%
\end{equation}
(and we also have the identity $\gamma_{k,+}^{2}-\gamma_{k,-}^{2}=4$). The $k$
dependence is dominated by the behaviour of $\gamma_{k,-}$. Now for large $k$%
\begin{equation}
\gamma_{k,-}\sim\frac{1}{2}\frac{m^{\left(  1\right)  2}-m^{\left(  2\right)
2}}{k^{2}}-\frac{1}{4k^{4}}\left(  m^{\left(  1\right)  4}-m^{\left(
2\right)  4}\right)  \label{expan}%
\end{equation}
and so there is a scale (determined by the ratio of the two terms in the
above)
\begin{equation}
k_{0}\sim\frac{1}{\sqrt{2}}\sqrt{\left(  m^{\left(  1\right)  2}+m^{\left(
2\right)  2}\right)  }~, \label{momcutoff}%
\end{equation}
which is a plausible cut-off scale in $k$. We note that $k_{0}\sim\overline
{m}+\frac{1}{8}$ $\frac{\delta m^{^{2}}}{\overline{m}}$ where $m^{\left(
1\right)  }=\overline{m}+\frac{1}{2}\delta m$, $m^{\left(  2\right)
}=\overline{m}-\frac{1}{2}\delta m$. Hence, there is a fall off for
$n_{\alpha,k}\left(  t\right)  $ with $k$ determined by $\gamma_{k,-}^{2}$ ,
which is qualitatively similar to the four-dimensional fermion case of
\cite{boyanovsky} and to an earlier conjectured cut-off $\frac{m_{1}+m_{2}}%
{2}$ \cite{barenboim} for $m_{1}\sim m_{2}$.

From these considerations, we obtain the flavour-vacuum contributions to the
cosmological constant by integrating over the extremely small time scales of
capture, which according to our previous discussion is qualitatively
equivalent to setting the time $\eta$-dependence to $\eta=0$. After taking
appropriately the continuum limit, $\frac{j}{L}\rightarrow\frac{k}{2\pi}$,
$\frac{1}{L}\sum_{j}\rightarrow\frac{1}{2\pi}\int dk\rightarrow\frac{1}{2\pi
}\int_{0}^{k_{0}}dk$, with the cutoff $k_{0}$ given by (\ref{momcutoff}), we
can obtain an estimate for the cosmological-constant contribution due to the
flavour-vacuum by considering the case of small expansion and expansion rate.
Returning to the expression for $\left.  _{in,\alpha,\beta}\right\langle
0\left\vert T_{00}^{out}\left\vert 0\right\rangle _{\alpha,\beta,in}\right.  $
in (\ref{energy}) we renormalise the vacuum expectation so as to leave the
contribution from the string capture of the stringy matter. One part of this
renormalisation is the C-independent subtraction which leads to
\begin{align}
&  \left.  _{in,\alpha,\beta}\right\langle 0\left\vert T_{00}^{out}\left\vert
0\right\rangle _{\alpha,\beta,in}\right. \nonumber\\
&  =\frac{C\left(  \eta\right)  }{8\pi}\sum_{i=1}^{2}\int_{0}^{k_{0}}%
dk\frac{m^{\left(  i\right)  2}}{\omega_{k}^{\left(  i\right)  }}\left(
\mathfrak{f}_{k,i}^{\left(  +\right)  }-\mathfrak{f}_{k,i}^{\left(  -\right)
}\right)  \label{subtraction1}%
\end{align}
and
\begin{align}
&  \left.  _{in,\alpha,\beta}\right\langle 0\left\vert T_{11}^{out}\left\vert
0\right\rangle _{\alpha,\beta,in}\right. \nonumber\\
&  =-\frac{C\left(  \eta\right)  }{8\pi}\sum_{i=1}^{2}\int_{0}^{k_{0}}%
dk\frac{m^{\left(  i\right)  2}}{\omega_{k}^{\left(  i\right)  }}\left(
\mathfrak{f}_{k,i}^{\left(  +\right)  }-\mathfrak{f}_{k,i}^{\left(  -\right)
}\right)  . \label{subtraction2}%
\end{align}
As noted earlier we already can see the D particle contribution to the flavour
vacuum giving $w=-1$. We further renormalise $\mathfrak{f}_{k,i}^{\left(
\pm\right)  }$ to only include the $\gamma_{k,-}^{in}$ dependent terms since
it is only such terms which distinguish it from the mass vacuum state. Details
are given in the Appendix . To leading order in $\delta m$, on averaging over
the uncertainty of the capture time $\eta$ (which is effectively
instantaneous) and owing to (\ref{fplusrel}) and (\ref{fminusrel})
\begin{align}
&  \left.  _{in,\alpha,\beta}\right\langle 0\left\vert T_{00}^{out}\left\vert
0\right\rangle _{\alpha,\beta,in}\right. \nonumber\\
&  =0.
\end{align}
In the next to leading order \ in $\delta m$%
\begin{align}
&  \left.  _{in,\alpha,\beta}\right\langle 0\left\vert T_{00}^{out}\left\vert
0\right\rangle _{\alpha,\beta,in}\right. \nonumber\\
&  \sim-\frac{\sin^{2}\theta}{24\pi}\left(  \delta m\right)  ^{2}%
+\frac{\left(  \delta m\right)  ^{2}}{8\pi}\sin^{2}\theta\,\left(  1-C\left(
\eta_{0}\right)  \right)
\end{align}
where $\eta_{0}$ is the capture time which in our model can be taken to be
$\eta_{0}=0$. For small $\widetilde{\sigma}$ we have $C\left(  \eta
_{0}\right)  \sim1-4\widetilde{\sigma}^{2}$ (c.f. (\ref{metric2b}),
(\ref{rwrecoil})). Hence, finally making the subtraction of the $\widetilde
{\sigma}$ independent terms we get
\begin{equation}
\left.  _{in,\alpha,\beta}\right\langle 0\left\vert \colon T_{00}^{out}%
\colon\left\vert 0\right\rangle _{\alpha,\beta,in}\right.  \sim\frac{\left(
\delta m\right)  ^{2}}{2\pi}\sin^{2}\theta\,\,\widetilde{\sigma}^{2}.
\label{flavourenergy}%
\end{equation}
This shows the contribution to the dark energy from the flavour-changing
D-brane-recoil (statistical) fluctuations.

The reader should notice that, within our effective field theory framework,
the (small) fluctuation parameter $\widetilde{\sigma}^{2}$ should be
considered as \emph{phenomenological}. In order to determine its real order of
magnitude, one needs detailed microscopic stringy models of such D-particle
foam, along the lines proposed in \cite{emw}, in the context of brane-world
cosmologies (c.f. fig.~\ref{fig:recoil}). This issue is still far from being
complete, given that realistic three-space-dimensional brane-world models of
D-particle foam require appropriate compactification, consistent with
phenomenologically realistic supersymmetry breaking scenarios. We postpone a
discussion of such issues for future work.

A related comment concerns the asymptotic in time nature of the
flavour-oscillation-induced de Sitter vacuum energy (\ref{flavourenergy}). The
reader should recall that this result was derived on the assumption that,
asymptotically in cosmic time, one has a uniform \emph{constant} density of
defects, and the space time approaches the flat, but not standard Minkowski
(i.e. $\widetilde{\sigma}^{2}$-dependent), limit (\ref{metric2b}). This will
cause eternal acceleration, as the situation is equivalent to that of a
cosmological constant, and perturbative Scattering matrix and asymptotic
states will not be well defined. This may have consequences on a quantum field
theory of matter defined in such space times, especially from the point of
view of an \emph{ill}-defined CPT operator, a situation which might arise in
such a case due to decoherence~\cite{wald} associated with the existence of
future horizons (cosmologically induced \emph{intrinsic} CPT
violation~\cite{mavromatos}).

On the other hand, in cases where the asymptotic density of defects, crossing
our brane world, vanishes, it is evident that the flavour-changing induced
vacuum energy (\ref{flavourenergy}) (and also pressure, in view of
(\ref{eos2}) will also \emph{vanish} asymptotically. In such a case, the
situation will resemble the one illustrated in the example of the space time
(\ref{spacetime2}) in section \ref{sec:expansion}, where the de Sitter phase
characterises only a certain era of the Universe, relaxing asymptotically to a
standard flat Minkowski space time. In the latter case, the perturbative
Scattering matrix and asymptotic states are then well defined and the
associated quantum field theory of matter fields on the brane world is CPT
invariant, at least from a cosmological view point.

\textbf{ }

\section{Conclusions and Outlook}

We have discussed here an approach to the flavour mixing problem and its
consequences to the vacuum energy in a toy model for bosons, taking into
account the effects of the Universe expansion.

We have attempted to argue that the flavour (Fock space) vacuum introduced by
Blasone and Vitiello in order to describe canonical quantization of theories
with mixing characterises certain string models entailing Lorentz violation,
in particular D-particle foam models. In such models, the interaction of
certain (electrically neutral) string excitations with the D-particle defects
cause back reaction effects onto the space time, due to the capture of the
string state by the D-particle and subsequent recoil of the space-time defect.
When a statistically significant population of such defects is present in our
brane world, then their effects are global, and may result in an expanding
Robertson-Walker Universe, depending on the details of the time profile of the
D-particle density. We have provided arguments in favour of such a behaviour
in the context of specific brane models propagating in a bulk space time,
punctured by D-particles.

During the capture/recoil process, flavour may not be conserved, and it is in
this sense that the D-foam background provide a way for flavour mixing. The
recoil of the defects cause Lorentz violation, as a result of the induced
momentum-dependent Finsler-type~\cite{finsler} metric in the process. Such
Lorentz violation may be averaged to zero in D-foam models, but leaves
traceable effects on quantum fluctuations.

In such an expanding Universe, with flavour mixing, the vacuum felt by the
string states is not the normal vacuum but the Fock-space flavoured vacuum
introduced by Blasone and Vitiello in theories with mixing. However, in
contrast to their generic considerations, in our approach the introduction of
the flavour vacuum is a result of a specific underlying microscopic model of
space-time background with defects interacting with some but not all matter excitations.

The details of the D-particle foam played an important r\^ole in determining
the characteristic time scales involved in the problem (especially the (small)
duration of the capture of the string state by the defect), and as a
consequence the pertinent time intervals over which one has to average in
order to obtain the contribution to the vacuum energy.

We have calculated the effects of the Universe expansion in the flavour-vacuum
formalism for a toy model of bosons. By taking into account the details of our
specific model involving D-particle defects, we have been able of determining
the physically relevant subtraction scheme in the calculation of the
flavour-vacuum expectation value of the matter stress-energy tensor. In this
way we have obtained an equation of state of cosmological-constant type for
(late) eras of the Universe, where the expansion rate is small. This is
consistent with the constant contributions to the vacuum energy obtained by
averaging over the relevant (small) time scales in the problem. Although our
considerations are toy, as referring to two-dimensional target space time, and
hence one cannot make direct contact with dark-energy observations,
nevertheless, the order of magnitude of this non perturbative contribution
(\ref{flavourenergy}) to the cosmological constant can be made
phenomenologically consistent with the small value suggested by observations
today, provided the order of magnitude of the recoil-velocity (statistical)
fluctuations $\sigma^{2}$, which it depends upon, is sufficiently small.
Moreover, the contribution vanishes if the mass difference or the mixing angle
vanish, as expected in the generic approach of Blasone and Vitiello.

We should extend the above analysis to include fermions, and also we should
consider higher-dimensional isotropic Universe situations. Moreover, despite
the motivations from string theory, our considerations have been restricted so
far in the context of effective low-energy field theories. It would be
interesting to extend these results to pure stringy considerations and attempt
to estimate microscopically the order of the recoil velocity fluctuations,
$\sigma^{2}$, in realistic models. Such a task is non trivial, as it requires
first an understanding of the details of the compactification procedure,
consistent with supersymmetry breaking in target space. Flux compactification
has been briefly used in our stringy model in the arguments in favour of the
identification of the Liouville mode of our supercritical recoil string model
with a function of target time. It is our belief that such compactifications
provide a way of resolving the problem of the smallness of the dark energy of
the Universe today consistently with phenomenologically realistic values of
supersymmetry-breaking mass scales. We hope to start tackling such issues in
the near future.

A final comment we would like to make concerns the global effect of the
induced de Sitter space time on photons. According to the analysis in
\cite{prokopec}, in four-dimensional de Sitter space times, vacuum
polarization of charged scalar fields, such as charged Higgs in supersymmetric
theories, implies mass for the long wavelength photon modes that cross the
horizon $H^{-1}$ of the (locally) de Sitter space time. Indeed, let $\vec k$
denote the momentum of such modes, with $c|\vec k | \gg H$ initially at some
(conformal) time $\eta_{0} = -H^{-1}$, but whose actual (physical) magnitude
is diminished significantly due to the Universe expansion at a later time
$\eta$, after a long period of inflation, $|\vec k |/a(\eta)$, with $a(\eta)$
the scale factor such that $a(t) \gg c|\vec k | \gg H$~. Then the induced mass
$m_{\gamma}$ of the photon is, to leading order in the appropriate
expansions~\cite{prokopec}:
\begin{equation}
m_{\gamma}\sim\sqrt{\alpha} \, c^{-2} \, \hbar\, H \, [\frac{2}{\pi
}\mathrm{ln}\left(  \frac{c\,|\vec k|}{H} \right)  + \mathcal{O}(1)]^{1/2}
\label{photonmass}%
\end{equation}
where $\alpha$ is the fine structure constant. In our case
(\ref{flavourenergy}), the flavour-vacuum-induced Hubble parameter is
extremely small (at late eras of the Universe), of order $|(\delta m )|
\mathrm{sin}^{\mathbf{1}/2}\theta\sqrt{\widetilde{\sigma}^{2}} $, and thus
only extremely long-wavelength (far infrared) photon modes would acquire a
(very small) mass via the vacuum-polarization mechanism of \cite{prokopec}.
However, in the early Universe, our statistical D-particle fluctuations could
be much larger, according to the models of D-foam discussed above, and thus
one does not exclude the possibility of significant photon masses (in
supersymmetric theories, where charged scalar fields occur naturally) via this
mechanism. As discussed in \cite{prokopec}, this may produce weak, but of
cosmological-scale relevance, \emph{seed} magnetic fields, which could be
amplified to produce the micro-Gauss magnetic fields observed in galaxies and
galactic clusters. It would be interesting to examine such scenarios in the
context of our flavour-vacuum induced de Sitter space time, as such studies
may provide indirect tests of (and impose constraints on) space time foam
models from novel view points. We hope to tackle such issues in a forthcoming publication.

\section*{Acknowledgements}

This work is partially supported by the European Union through the FP6 Marie
Curie Research and Training Network \emph{UniverseNet} (MRTN-CT-2006-035863).

\section{\textbf{Appendix }}

In this Appendix we calculate the quantities $\mathfrak{f}_{k,i}^{\left(
\pm\right)  } $, appearing in the appropriately normal-ordered (subtracted)
expressions (\ref{subtraction1}), (\ref{subtraction2}) for the boson
stress-energy tensor flavour-vacuum expectation value, for slow
Universe-expansion rate and small momenta $k$ (with respect to the boson mass).

It is straightforward to demonstrate that%
\begin{align}
\mathfrak{f}_{k,i}^{\left(  +\right)  }  &  =-\left[  \left\vert
\mathfrak{A}_{k,i}^{\left(  1\right)  }\right\vert ^{2}+\left\vert
\mathfrak{A}_{k,i}^{\left(  2\right)  }\right\vert ^{2}+\left\vert
\widetilde{\mathfrak{A}}_{-k,i}^{\left(  1\right)  }\right\vert ^{2}%
+\left\vert \widetilde{\mathfrak{A}}_{-k,i}^{\left(  2\right)  }\right\vert
^{2}\right] \label{matrixel3}\\
\mathfrak{f}_{k,i}^{\left(  -\right)  }  &  =-\left[  \mathfrak{A}%
_{k,i}^{\left(  1\right)  }\widetilde{\mathfrak{A}}_{k,i}^{\left(  1\right)
\ast}+\mathfrak{A}_{k,i}^{\left(  2\right)  }\widetilde{\mathfrak{A}}%
_{k,i}^{\left(  2\right)  \ast}+\widetilde{\mathfrak{A}}_{-k,i}^{\left(
1\right)  }\mathfrak{A}_{-k,i}^{\left(  1\right)  \ast}+\widetilde
{\mathfrak{A}}_{-k,i}^{\left(  2\right)  }\mathfrak{A}_{-k,i}^{\left(
2\right)  \ast}\right]  \label{matrixel4}%
\end{align}
where
\begin{align}
\mathfrak{A}_{k,i}^{\left(  1\right)  }  &  =-\delta_{i1}\,\alpha_{k}^{\left(
1\right)  }e^{i\left(  kx-\omega_{k}^{\left(  1\right)  out}\eta\right)  }%
\cos\theta\,\nonumber\\
&  +\frac{1}{2}\delta_{i2}\,e^{i\left(  kx-\omega_{k}^{\left(  2\right)
out}\eta\right)  }\left[
\begin{array}
[c]{c}%
-\alpha_{k}^{\left(  2\right)  }\gamma_{k,+}^{in}e^{i\left(  \omega
_{k}^{\left(  2\right)  in}-\omega_{k}^{\left(  1\right)  in}\right)  \eta}\\
+\beta_{k}^{\left(  2\right)  \ast}\gamma_{k,-}^{in}e^{-i\left(  \omega
_{k}^{\left(  2\right)  in}+\omega_{k}^{\left(  1\right)  in}\right)  \eta}%
\end{array}
\right]  \sin\theta, \label{A1}%
\end{align}%
\begin{align}
\mathfrak{A}_{k,i}^{\left(  2\right)  }  &  =\frac{1}{2}\delta_{i1}e^{i\left(
kx-\omega_{k}^{\left(  1\right)  out}\eta\right)  }\sin\theta\left\{
\gamma_{k,+}^{in}\alpha_{k}^{\left(  1\right)  }e^{-i\left(  \omega
_{k}^{\left(  2\right)  in}-\omega_{k}^{\left(  1\right)  in}\right)  \eta
}+\gamma_{k,-}^{in}\beta_{k}^{\left(  1\right)  \ast}e^{-i\left(  \omega
_{k}^{\left(  2\right)  in}+\omega_{k}^{\left(  1\right)  in}\right)  \eta
}\right\} \nonumber\\
&  -\delta_{i2}\,e^{i\left(  kx-\omega_{k}^{\left(  2\right)  out}\eta\right)
}\alpha_{k}^{\left(  2\right)  }\cos\theta\;, \label{A2}%
\end{align}

\begin{align}
\widetilde{\mathfrak{A}}_{-k,i}^{\left(  1\right)  }  &  =\delta
_{i1}\,e^{-i\left(  kx-\omega_{k}^{\left(  1\right)  out}\eta\right)  }%
\beta_{j}^{\left(  1\right)  }\cos\theta\nonumber\\
&  +\frac{1}{2}\delta_{i2}\,e^{-i\left(  kx-\omega_{k}^{\left(  2\right)
out}\eta\right)  }\sin\theta\nonumber\\
&  \times\left\{  \beta_{k}^{\left(  2\right)  }\gamma_{k,+}^{in}e^{-i\left(
-\omega_{k}^{\left(  2\right)  in}+\omega_{k}^{\left(  1\right)  in}\right)
\eta}-\alpha_{k}^{\left(  2\right)  \ast}\gamma_{k,-}^{in}e^{-i\left(
\omega_{k}^{\left(  2\right)  in}+\omega_{k}^{\left(  1\right)  in}\right)
\eta}\right\}  , \label{AT1}%
\end{align}
and
\begin{align}
\widetilde{\mathfrak{A}}_{-k,i}^{\left(  2\right)  }  &  =-\frac{1}{2}%
\delta_{i1}\,e^{-i\left(  kx-\omega_{k}^{\left(  1\right)  out}\eta\right)
}\sin\theta\nonumber\\
&  \times\left\{  \alpha_{k}^{\left(  1\right)  \ast}\gamma_{k,-}%
^{in}\,e^{-i\left(  \omega_{k}^{\left(  2\right)  in}+\omega_{k}^{\left(
1\right)  in}\right)  \eta}+\beta_{k}^{\left(  1\right)  }\gamma_{k,+}%
^{in}\,e^{-i\left(  \omega_{k}^{\left(  2\right)  in}-\omega_{k}^{\left(
1\right)  in}\right)  \eta}\right\} \nonumber\\
&  +\delta_{i2}\,e^{-i\left(  kx-\omega_{k}^{\left(  2\right)  out}%
\eta\right)  }\beta_{k}^{\left(  2\right)  }\cos\theta. \label{AT2}%
\end{align}
We note the general identity $\gamma_{k,+}^{2}=4+\gamma_{k,-}^{2}$ and that,
for small $k$,
\begin{equation}
\gamma_{k,-}^{in}\simeq\frac{\delta m}{\sqrt{m^{\left(  1\right)  }m^{\left(
2\right)  }}}-\frac{2\Delta m^{2}\,m}{4\lambda^{2}\left(  m^{\left(  1\right)
}m^{\left(  2\right)  }\right)  ^{\frac{5}{2}}}k^{2} \label{expn2}%
\end{equation}
where $\ A=\frac{1+\lambda^{2}}{2},\,B=\frac{1-\lambda^{2}}{2}$and $\Delta
m^{2}\equiv m^{\left(  1\right)  2}-m^{\left(  2\right)  2}$. In terms of the
D-particle model $\lambda=1-4\sigma^{2}$ when $\sigma$ is small. The \emph{in}
flavour vacuum differs from the mass eigenstate vacuum through terms
proportional to $\gamma_{k,-}^{in}$ and so the effect of the D-particle
capture which we regard as responsible for the mixing in this model is present
through such terms in $\mathfrak{f}_{k,i}^{\left(  +\right)  }$ and
$\mathfrak{f}_{k,i}^{\left(  -\right)  }$. Consequently we write for $s=1,2$
\begin{equation}
\mathfrak{A}_{k,i}^{\left(  s\right)  }=\mathfrak{g}_{k,i}^{\left(
s,0\right)  }+\mathfrak{g}_{k,i}^{\left(  s,1\right)  }\gamma_{k,-}%
^{in}+O\left(  \gamma_{k,-}^{in2}\right)  \label{AS}%
\end{equation}
and
\begin{equation}
\widetilde{\mathfrak{A}}_{-k,i}^{\left(  s\right)  }=\widetilde{\mathfrak{g}%
}_{-k,i}^{\left(  s,0\right)  }+\widetilde{\mathfrak{g}}_{k,i}^{\left(
s,1\right)  }\gamma_{k,-}^{in}+O\left(  \gamma_{k,-}^{in2}\right)  \label{ATS}%
\end{equation}
where
\begin{align}
\mathfrak{g}_{k,i}^{\left(  s,0\right)  }  &  =-\delta_{is}\alpha_{k}^{\left(
s\right)  }e^{i\left(  kx-\omega_{k}^{\left(  s\right)  out}\eta\right)  }%
\cos\theta\,\nonumber\\
&  +\left(  -1\right)  ^{s}\delta_{i\left[  s+1\right]  _{2}}\alpha
_{k}^{\left(  \left[  s+1\right]  _{2}\right)  }e^{i\left(  kx-\left(
\omega_{k}^{\left(  \left[  s+1\right]  _{2}\right)  out}-\omega_{k}^{\left(
\left[  s+1\right]  _{2}\right)  in}+\omega_{k}^{\left(  s\right)  in}\right)
\eta\right)  }\sin\theta,
\end{align}%
\begin{equation}
\mathfrak{g}_{k,i}^{\left(  s,1\right)  }=\frac{1}{2}\delta_{i\left[
s+1\right]  _{2}}e^{i\left(  kx-\left(  \omega_{k}^{\left(  \left[
s+1\right]  _{2}\right)  out}+\omega_{k}^{\left(  \left[  s+1\right]
_{2}\right)  in}+\omega_{k}^{\left(  s\right)  in}\right)  \eta\right)  }%
\beta_{k}^{\left(  \left[  s+1\right]  _{2}\right)  \ast},
\end{equation}%
\begin{align}
\widetilde{\mathfrak{g}}_{-k,i}^{\left(  s,0\right)  }  &  =\delta
_{is}e^{-i\left(  kx-\omega_{k}^{\left(  s\right)  out}\eta\right)  }\beta
_{k}^{\left(  s\right)  }\cos\theta\\
&  -\left(  -1\right)  ^{s}\delta_{i\left[  s+1\right]  _{2}}e^{-i\left(
kx-\left(  \omega_{k}^{\left(  \left[  s+1\right]  _{2}\right)  out}%
+\omega_{k}^{\left(  \left[  s+1\right]  _{2}\right)  in}-\omega_{k}^{\left(
s\right)  in}\right)  \eta\right)  }\beta_{k}^{\left(  \left[  s+1\right]
_{2}\right)  }\sin\theta,\nonumber
\end{align}
and
\begin{equation}
\widetilde{\mathfrak{g}}_{-k,i}^{\left(  s,1\right)  }=-\frac{1}{2}%
\delta_{i\left[  s+1\right]  _{2}}\alpha_{k}^{\left(  \left[  s+1\right]
_{2}\right)  \ast}e^{-i\left(  kx-\left(  \omega_{k}^{\left(  \left[
s+1\right]  _{2}\right)  out}-\omega_{k}^{\left(  \left[  s+1\right]
_{2}\right)  in}-\omega_{k}^{\left(  s\right)  in}\right)  \eta\right)  }%
\sin\theta
\end{equation}
with $\left[  1\right]  _{2}=1,\left[  2\right]  _{2}=2$ and $\left[
3\right]  _{2}=1$. The leading contributions to the renormalised
$\colon\mathfrak{f}_{k,i}^{\left(  -\right)  }\colon$ and $\colon
\mathfrak{f}_{k,i}^{\left(  +\right)  }\colon$ are%
\begin{equation}
\colon\mathfrak{f}_{k,i}^{\left(  -\right)  }\colon=-\sum_{s=1}^{2}\left(
\mathfrak{g}_{k,i}^{\left(  s,1\right)  }\widetilde{\mathfrak{g}}%
_{k,i}^{\left(  s,0\right)  \ast}+\mathfrak{g}_{k,i}^{\left(  s,0\right)
}\widetilde{\mathfrak{g}}_{k,i}^{\left(  s,1\right)  \ast}+\mathfrak{g}%
_{-k,i}^{\left(  s,0\right)  \ast}\widetilde{\mathfrak{g}}_{-k,i}^{\left(
s,1\right)  }+\mathfrak{g}_{-k,i}^{\left(  s,1\right)  \ast}\widetilde
{\mathfrak{g}}_{-k,i}^{\left(  s,0\right)  }\right)  \gamma_{k,-}^{in}
\label{renormminus}%
\end{equation}
and
\begin{equation}
\colon\mathfrak{f}_{k,i}^{\left(  +\right)  }\colon=-\sum_{s=1}^{2}\left(
\mathfrak{g}_{k,i}^{\left(  s,0\right)  }\mathfrak{g}_{k,i}^{\left(
s,1\right)  \ast}+\mathfrak{g}_{k,i}^{\left(  s,1\right)  }\mathfrak{g}%
_{k,i}^{\left(  s,0\right)  \ast}+\widetilde{\mathfrak{g}}_{-k,i}^{\left(
s,0\right)  }\widetilde{\mathfrak{g}}_{-k,i}^{\left(  s,1\right)  \ast
}+\widetilde{\mathfrak{g}}_{-k,i}^{\left(  s,1\right)  }\widetilde
{\mathfrak{g}}_{-k,i}^{\left(  s,0\right)  \ast}\right)  \gamma_{k,-}^{in}.
\label{renormplus}%
\end{equation}
Now on noting, for real $y$, that $\Gamma\left(  iy\right)  \Gamma\left(
-iy\right)  =\frac{\pi}{y\sinh\left(  \pi y\right)  }$ and $\arg\Gamma\left(
iy\right)  \sim y\log y-y-\frac{\pi}{4}+O\left(  \frac{1}{y}\right)  $ as
$y\rightarrow\infty$ we can show that for small $\left\vert 1-\lambda
\right\vert $ and $\rho$ that
\begin{equation}
\alpha_{k}^{\left(  i\right)  }\sim\exp\left[  -\frac{i}{4\rho}\frac
{m^{\left(  i\right)  4}\left(  \lambda-1\right)  ^{2}}{\left(  k^{2}%
+m^{\left(  i\right)  2}\right)  ^{3/2}}\right]
\end{equation}
and
\begin{equation}
\beta_{k}^{\left(  i\right)  }\sim-i\exp\left[  -\frac{\pi}{\rho}\left(
\lambda m^{\left(  i\right)  }+\frac{k^{2}}{2\lambda m^{\left(  i\right)  }%
}\right)  \right]  .
\end{equation}
This leads to
\begin{align}
&  \colon\mathfrak{f}_{k,i}^{\left(  +\right)  }\colon\nonumber\\
&  =2\gamma_{k,-}^{in}\sin^{2}\theta\,\sum_{s=1}^{2}\left(  -1\right)
^{s}\delta_{i\left[  s+1\right]  _{2}}\exp\left(  -\frac{\pi}{\rho}\left[
\lambda m^{\left(  \left[  s+1\right]  _{2}\right)  }+\frac{k^{2}}{2\lambda
m^{\left(  \left[  s+1\right]  _{2}\right)  }}\right]  \right) \nonumber\\
&  \times\sin\left(  \frac{m^{\left(  \left[  s+1\right]  _{2}\right)
4}\left(  \lambda-1\right)  ^{2}}{4\rho\left(  k^{2}+m^{\left(  \left[
s+1\right]  _{2}\right)  2}\right)  ^{\frac{3}{2}}}\right)
\label{fsimpleplus}%
\end{align}
and
\begin{align}
&  \colon\mathfrak{f}_{k,i}^{\left(  -\right)  }\colon\nonumber\\
&  =-\gamma_{k,-}^{in}\sin^{2}\theta\sum_{s=1}^{2}\left(  -1\right)
^{s+1}\delta_{i\left[  s+1\right]  _{2}}\left\{
\begin{array}
[c]{c}%
\cos\left(  \frac{m^{\left(  \left[  s+1\right]  _{2}\right)  4}\left(
\lambda-1\right)  ^{2}}{2\rho\left(  k^{2}+m^{\left(  \left[  s+1\right]
_{2}\right)  2}\right)  ^{\frac{3}{2}}}\right) \\
-\exp\left(  -\frac{2\pi}{\rho}\left(  \lambda m^{\left(  \left[  s+1\right]
_{2}\right)  }+\frac{k^{2}}{2\lambda m^{\left(  \left[  s+1\right]
_{2}\right)  }}\right)  \right)
\end{array}
\right\}  . \label{fsimpleminus}%
\end{align}
To lowest order in $\delta m$
\begin{equation}
\colon\mathfrak{f}_{k,2}^{\left(  +\right)  }\colon=-\colon\mathfrak{f}%
_{k,1}^{\left(  +\right)  }\colon, \label{fplusrel}%
\end{equation}%
\begin{equation}
\colon\mathfrak{f}_{k,2}^{\left(  -\right)  }\colon=-\colon\mathfrak{f}%
_{k,1}^{\left(  -\right)  }\colon, \label{fminusrel}%
\end{equation}
and (c.f. (\ref{gplusminus}))
\begin{equation}
\gamma_{k,-}^{in}=\frac{\delta m}{m}~, \label{gsimpminus}%
\end{equation}
which we use in (\ref{flavourenergy}) to estimate the leading contribution to
the dark energy coming from the flavour-changing D-brane-reoil fluctuations in
our model.

\bigskip

\bigskip\

\end{document}